\documentclass[english,showkeys,notitlepage]{revtex4}
\usepackage[T1]{fontenc}
\usepackage[latin9]{inputenc}
\usepackage{color}
\usepackage{babel}
\usepackage{amsmath}
\usepackage{amssymb}
\usepackage{graphicx}
\usepackage{esint}
\usepackage[unicode=true,pdfusetitle,
 bookmarks=true,bookmarksnumbered=false,bookmarksopen=false,
 breaklinks=false,pdfborder={0 0 1},backref=false,colorlinks=false]
 {hyperref}

\makeatletter

\providecommand{\tabularnewline}{\\}

\@ifundefined{textcolor}{}
{%
 \definecolor{BLACK}{gray}{0}
 \definecolor{WHITE}{gray}{1}
 \definecolor{RED}{rgb}{1,0,0}
 \definecolor{GREEN}{rgb}{0,1,0}
 \definecolor{BLUE}{rgb}{0,0,1}
 \definecolor{CYAN}{cmyk}{1,0,0,0}
 \definecolor{MAGENTA}{cmyk}{0,1,0,0}
 \definecolor{YELLOW}{cmyk}{0,0,1,0}
}

\providecommand{\tabularnewline}{\\}

\makeatother

\begin{document}

\title{Single diffractive production of open heavy flavor mesons}

\author{Marat Siddikov, Iván Schmidt}

\affiliation{Departamento de Física, Universidad Técnica Federico Santa María,~~\\
 y Centro Científico - Tecnológico de Valparaíso, Casilla 110-V, Valparaíso,
Chile}
\begin{abstract}
In this paper we discuss the single diffractive production of open
heavy flavor mesons and non-prompt charmonia in $pp$ collisions.
Using the color dipole approach, we found that the single diffractive
production constitutes 0.5-2 per cent of the inclusive production
of the same mesons. In Tevatron kinematics our theoretical results
are in reasonable agreement with the available experimental data.
In LHC kinematics we found that the cross-section is sufficiently
large and could be accessed experimentally. We also analyzed the dependence
on multiplicity of co-produced hadrons and found that it is significantly
slower than that of inclusive production of the same heavy mesons. 
\end{abstract}
\maketitle

\section{Introduction}

In the kinematics of the Large Hadronic Collider (LHC), the diffractive
events in $pp$ collisions constitute approximately twenty per cent
of all inclusive events~\cite{Abelev:2012sea}, and for this reason
might be used as an additional tool for studies of the strong interactions.
The characteristic feature of the diffractive events is the presence
of rapidity gaps between hadronic products in the final state. In
Quantum Chromodynamics (QCD) such rapidity gaps in high energy kinematics
are explained by the exchange of pomerons in the $t$-channel. Since
the structure of the pomeron is relatively well understood and largely
does not depend on the process, the existence of rapidity gap allows
to separate the strong interactions involving different hadrons. While
conventionally diffractive production of mesons has been studied in
$ep$ collisions, there are various theoretical suggestions to use
$pp$ collisions for studies of the diffractive production of prompt
quarkonia~\cite{Cisek:2016kvr,Machado:2008zza,Machado:2007vw,Yuan:1998rq,Yuan:1998qw},
dijets~\cite{Mantysaari:2019csc}, gauge bosons~\cite{Pasechnik:2012ac},
Higgs bosons~\cite{Pasechnik:2014lga}, heavy quarks~\cite{Kopeliovich:2007vs,Luszczak:2014cxa},
quarkonia pairs~\cite{BrennerMariotto:2018tpq} and Drell-Yan processes~\cite{Kopeliovich:2006tk}.
The possibility to measure diffractive production in $pp$ collisions
has been demonstrated at the Tevatron~\cite{Affolder:2001nc,Affolder:1999hm,Aaltonen:2012tha,Aaltonen:2010qe,Affolder:2001zn}.
At the LHC some diffractive processes (\emph{e.g}. single diffractive
$pp\to pX$) have been measured with very good precision~\cite{Abelev:2012sea},
although diffractive production of additional heavy hadrons so far
has not been explored in depth (see however preliminary feasibility
study~\cite{CMS:2014rga}).

In this paper we are going to focus on single diffractive production
of heavy mesons, $pp\to p\,+M\,X$, where $M$ is an open heavy flavor
meson ($D$ or $B$) or a charmonium produced from decay of $B$-meson;
we also assume that the recoil proton in the final state is separated
by a rapidity gap from other hadrons. This process deserves special
interest both on its own and because it could help to clarify the
role of multipomeron contributions to the production of heavy quarks
in general. The role of such mechanisms is not very clear at this
moment. Usually it is believed that production of heavy quarks might
be described perturbatively~\cite{Korner:1991kf,Neubert:1993mb}
and is dominated by two-gluon (pomeron) fusion~\cite{Bodwin:1994jh,Maltoni:1997pt,Binnewies:1998vm,Kniehl:1999vf,Ma:2018bax,Goncalves:2017chx,Brambilla:2008zg,Feng:2015cba,Brambilla:2010cs}.
However, this approach can hardly explain the recently measured dependence
of the production cross-sections on the multiplicity of the charged
hadrons co-produced together with a given heavy quarkonia~\cite{Adam:2015ota,Trzeciak:2015fgz,Ma:2016djk,PSIMULT,Khatun:2019slm,Alice:2012Mult}.
Potentially this discrepancy might indicate sizeable contributions
of multigluon production mechanisms. At the same time, for $D$- and
$B$-mesons such rapidly growing dependence was not observed~\cite{Adam:2015ota}.
On the other hand, theoretical studies~\cite{Motyka:2015kta,Levin:2018qxa,Siddikov:2019xvf}
found that three-pomeron mechanism might give sizeable contribution
and can explain the observed multiplicity dependence of quarkonia.
For $D$-mesons it was found in the same framework that the
three-pomeron correction is also pronounced and might constitute up
to 40 percent of the result, although in the range of multiplicities
available at present from the LHC it does not contribute to the observed
multiplicity dependence due to partial cancellation with certain interference
contributions~\cite{Schmidt:2020fgn}. Fortunately, it is possible
to estimate the role of the three-pomeron fusion directly. The single
diffractive production at the partonic level has a similar structure,
and thus might provide independent estimate of the three-pomeron contribution.
Since the single diffractive production amplitude includes only one
cut pomeron which might contribute to the observed yields of co-produced
hadrons, its cross-section might be used as a very clean probe of
the multiplicity dependence of individual cut pomerons in high multiplicity
events.

Earlier the single-diffractive production including heavy quarks has
been studied in~~\cite{Kopeliovich:2007vs,Luszczak:2014cxa} for
the case of prompt production of quarkonia. As we will see below,
the cross-sections of single diffractive production of $D$- and $B$-mesons
is larger than that of the prompt charmonia and thus could be easier
to study experimentally. The feasibility to measure such processes
has been discussed in~\cite{CMS:2014rga,Affolder:2001nc,Affolder:1999hm}.
The study of rare events with large multiplicity requires
better statistics, and for this reason we expect that such dependence
could be measured during the High Luminosity Run 3 at the LHC (HL-LHC
mode)~\cite{ATLAS:2013hta,Apollinari:HLLHC,LaRoccaRiggi}.

The paper is structured as follows. In Section~\ref{sec:Evaluation}
we develop the general framework for the evaluation of the open heavy
meson production. We will perform our calculations within the color
dipole framework, which describes correctly the onset of saturation
dynamics and thus might be used even for the description of high multiplicity
events. In Section~\ref{sec:Numer} we present our numerical results
and make comparison with experimental data available from the Tevatron,
as well as with other theoretical approaches. In Section~\ref{sec:MultiplicityGeneralities}
we develop the framework for the description of multiplicity dependence
in dipole framework and compare its predictions for multiplicity dependence
with that of inclusive production. In Section~\ref{sec:Nucl} we
discuss briefly the single diffractive process on nuclei, $pA\to p\,+MX$.
Finally, in Section~\ref{sec:Conclusions} we draw conclusions.

\section{Single-Diffractive Production in color dipole framework}

\label{sec:Evaluation}
\begin{figure}
\includegraphics[width=9cm]{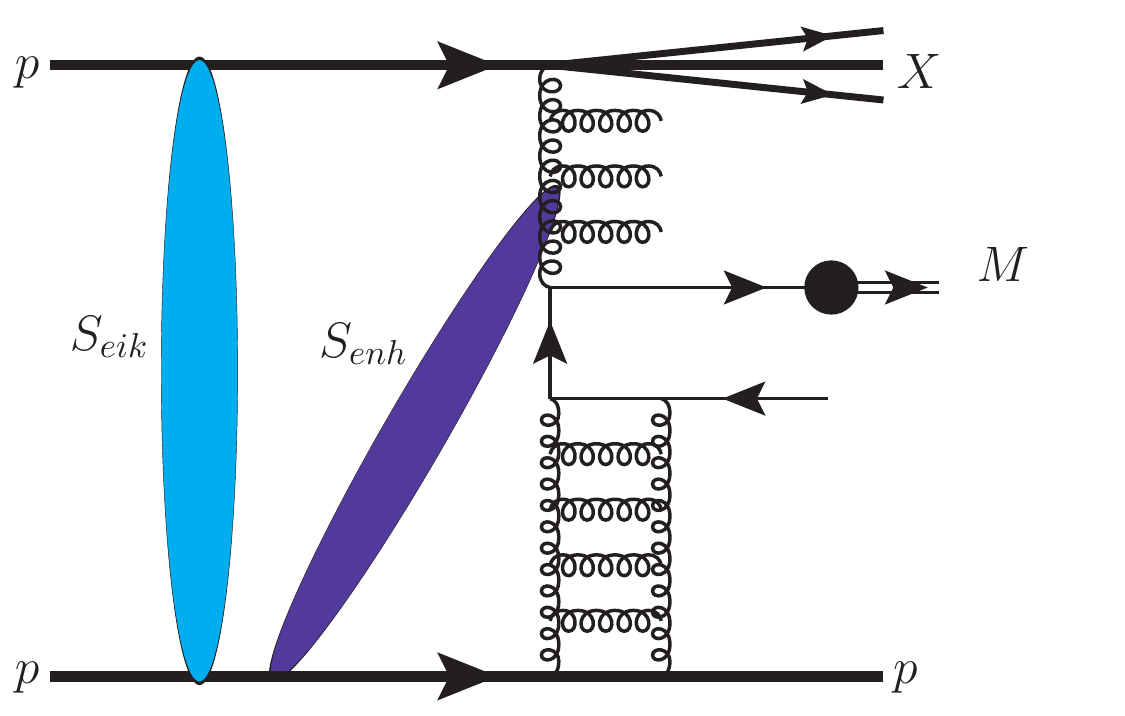}\includegraphics[width=9cm]{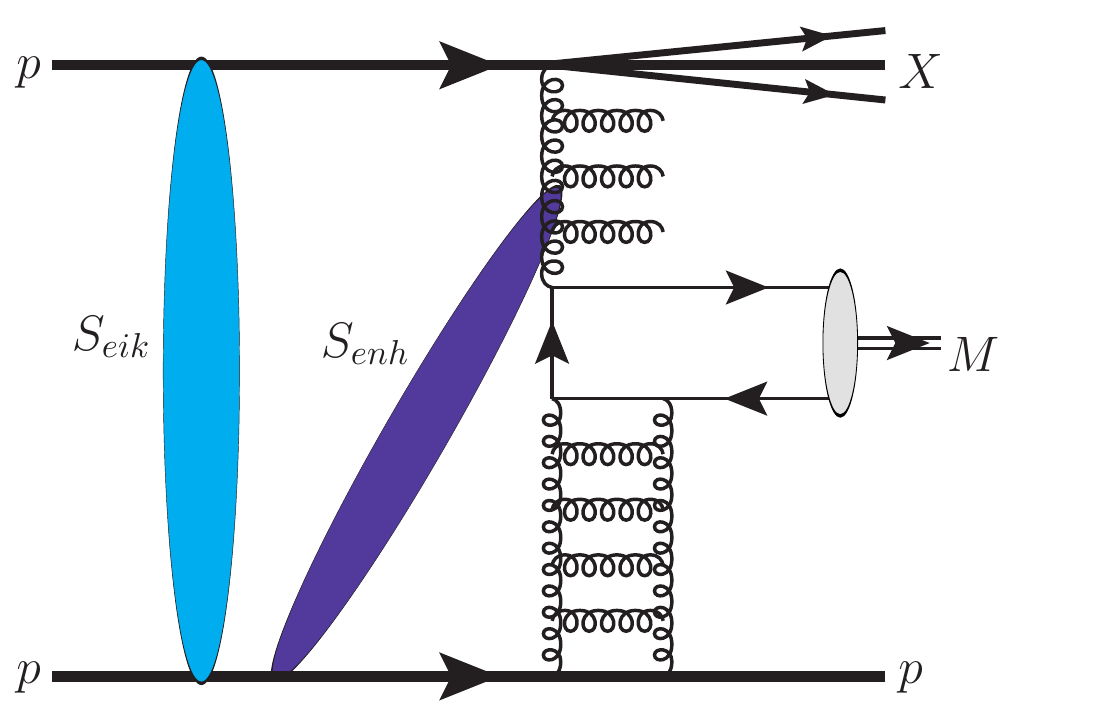}

\caption{\label{fig:SD}Left plot: The leading order contribution to single
diffractive production of open heavy flavor quark mesons. The recoil
proton (lower part) is separated from the heavy hadron by a rapidity
gap. The colored vertical and inclined ovals schematically illustrate
the contributions of the secondary interactions, whose products might
fill the rapidity gap between the recoil proton and the other hadrons
(see the text for discussion). Right plot: The leading order contribution
to the single diffractive production of prompt charmonia studied in~\cite{Machado:2007vw,Yuan:1998rq,Yuan:1998qw}.}
\end{figure}

As was mentioned in the previous section, a
defining characteristics of the single-diffractive production is the
observation of the recoil proton separated by a large rapidity gap
from other hadrons. In LHC kinematics the dominant contribution to
such process stems from the diagrams which include the exchange of
uncut pomeron between the proton and the other hadrons in the $t$-channel.
The heavy mesons are produced predominantly near the edge of the rapidity
gap, and for this reason a pomeron couples directly to the heavy quark
loop, as shown in the Figure~\ref{fig:SD}. In this paper we will
focus on the production of open heavy-flavor $D$- and $B$-mesons,
and will also discuss briefly the production of non-prompt charmonia
from decays of $B$-meson. Previously, the single diffractive production
for \emph{prompt} charmonia production has been studied in~\cite{Machado:2007vw,Yuan:1998rq,Yuan:1998qw}.
In this last case the dominant contribution differs slightly from
that of $D$- and $B$-mesons and is shown in the right panel of the
Figure~\ref{fig:SD}. In Section~\ref{sec:Numer} we will use the
results of~\cite{Machado:2007vw,Yuan:1998rq,Yuan:1998qw} for comparison
with our numerical results for non-prompt charmonia.

The cross-section of the heavy meson production might be related to
the cross-section of the heavy quark production as~\cite{Binnewies:1998vm,Kniehl:1999vf,Ma:2018bax,Goncalves:2017chx}.
\begin{equation}
\frac{d\sigma_{M}}{dy\,d^{2}p_{T}}=\sum_{i}\int_{x_{Q}}^{1}\frac{dz}{z^{2}}D_{i}\left(\frac{x_{Q}(y)}{z}\right)\,\frac{d\sigma_{\bar{Q}_{i}Q_{i}}}{dy^{*}d^{2}p_{T}^{*}}\label{eq:fragConvolution}
\end{equation}
where $y$ is the rapidity of the heavy meson ($D$- or $B$-meson),
$y^{*}=y-\ln z$ is the rapidity of the heavy quark, $p_{T}$ is the
transverse momentum of the produced $D$-meson, $D_{i}(z)$ is the
fragmentation function, which describes the parton $i$ fragmentation
into a heavy meson, and $d\sigma_{\bar{Q}_{i}Q_{i}}$ is the cross-section
of a heavy quark production with a rapidity $y^{*}$, discussed below
in Subsection~\ref{subsec:3Pom-1}. The dominant contribution to
all heavy mesons stems from the $c$- and $b$-quarks (prompt and
non-prompt mechanisms respectively), so the $d\sigma_{\bar{Q}_{i}Q_{i}}$
might be evaluated in the heavy quark mass limit. The fragmentation
functions for the $D$- and $B$-mesons, as well as non-prompt $J/\psi$
production, are known from the literature and for the sake of completeness
are given in Appendix~\ref{sec:FragFunctions}.

In Figure~\ref{fig:SD} we also included colored oval blobs, which
stand schematically for the secondary interactions which potentially
could fill the large rapidity gap in the final state. The general
framework for the evaluation of the rapidity gap survival factors
(\emph{i.e}. the probability that no particles will be produced in
a rapidity gap) has been developed in~\cite{Martin:2008nx,Khoze:2018kna,Ryskin:2009tj,Khoze:2000,Khoze:2017sdd},
and is briefly discussed below in Section~\ref{subsec:LRGSF}.

\subsection{Leading order single diffractive contribution}

\label{subsec:3Pom-1} 
\begin{figure}
\includegraphics[width=9cm]{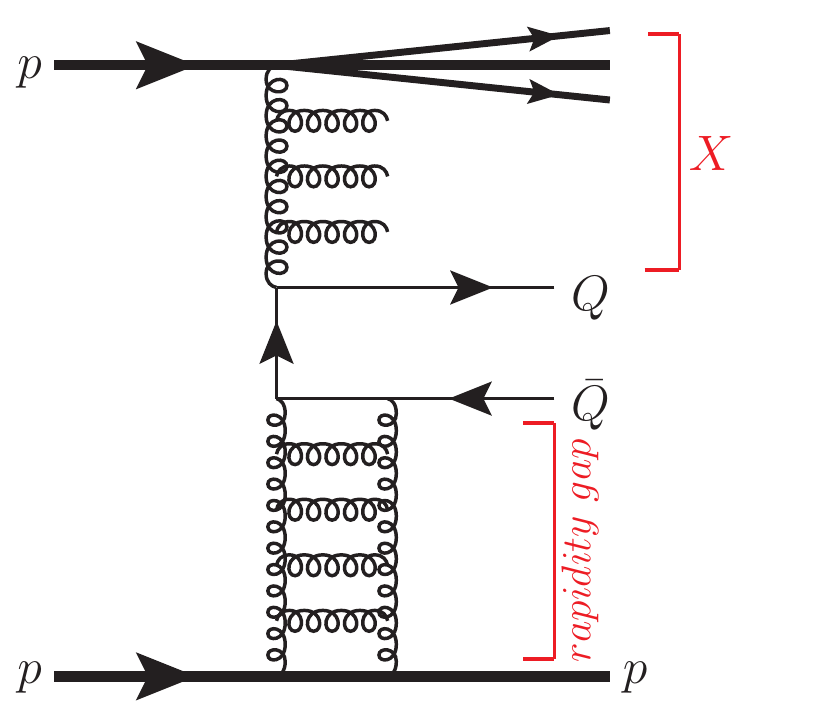}\includegraphics[width=9cm]{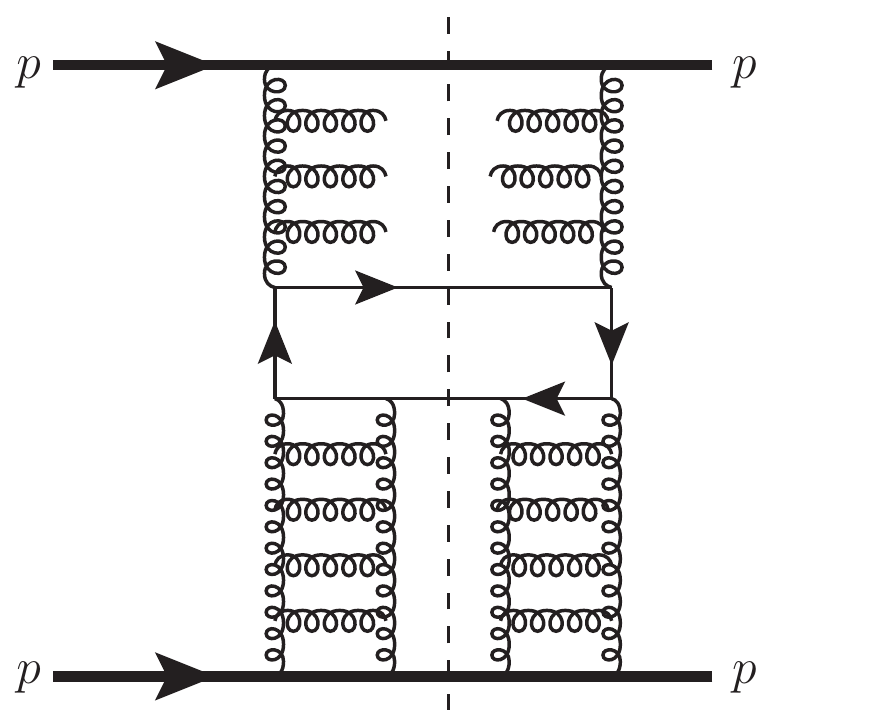}\caption{\label{fig:DipoleCrossSections-2Pom}Left plot: The leading order
contribution to the amplitude of single diffractive production of
heavy quarks separated by a rapidity gap from the recoil proton. Right
plot: Illustration indicating how the cross-section of the process
is related to the production amplitude from three pomeron fusion.
The dashed vertical line stands for the unitarity cut. The diagram
includes one cut pomeron (upper gluon ladder) and two uncut pomerons
(lower gluon ladders). In both plots a summation over all possible
permutations of gluon vertices in the heavy quark line/loop is implied.}
\end{figure}

The single diffractive production of onshell heavy quark pair in the
reference frame of the recoil proton might be viewed as a fluctutation
of the incoming virtual gluon into a heavy $\bar{Q}Q$ pair, with
subsequent \emph{elastic} scattering of the $\bar{Q}Q$ dipole on
the target proton. In perturbative QCD the dominant contribution to
such process is given by the diagram which includes exchange of a
single pomeron between $Q\bar{Q}$ and a recoil proton, in the spirit
of the Ingelman-Schlein model~\cite{Ingelman:1984ns} (see Figure~~\ref{fig:DipoleCrossSections-2Pom}
for details). In LHC kinematics the typical light-cone momentum fractions
$x_{1,2}$ carried by gluons are very small ($\ll1$), so the gluon
densities are enhanced in this kinematics. This enhancement modifies
some expectations based on the heavy quark mass limit. For example,
there could be sizeable corrections from multiple pomeron exchanges
between the heavy dipole and the target. For this reason instead of
hard process on individual partons it is more appropriate to use the
color dipole framework (also known as CGC/Sat)~\cite{GLR,McLerran:1993ni,McLerran:1993ka,McLerran:1994vd,MUQI,MV,gbw01:1,Kopeliovich:2002yv,Kopeliovich:2001ee}.
At high energies the color dipoles are eigenstates of interaction,
and thus can be used as the universal elementary building blocks automatically
accumulating both the hard and soft fluctuations~\cite{Nikolaev:1994kk}.
The light-cone color dipole framework has been developed and successfully
applied to phenomenological description of both hadron-hadron and
lepton-hadron collisions~\cite{Kovchegov:1999yj,Kovchegov:2006vj,Balitsky:2008zza,Kovchegov:2012mbw,Balitsky:2001re,Cougoulic:2019aja,Aidala:2020mzt,Ma:2014mri}.
Another advantage of the CGC/Sat (color dipole) framework is that
it allows a relatively straightforward extension for the description
of high-multiplicity events, as discussed in~\cite{KOLEB,KLN,DKLN,Kharzeev:2000ph,Kovchegov:2000hz,LERE,Lappi:2011gu,Ma:2018bax}.
The cross-section of the single diffractive process, shown in Figure~\ref{fig:DipoleCrossSections-2Pom},
in the dipole approach is given by 
\begin{eqnarray}
 &  & \frac{d\sigma_{\bar{Q}_{i}Q_{i}}\left(y,\,\sqrt{s}\right)}{dy\,d^{2}p_{T}}=\,\int d^{2}k_{T}x_{1}\,g\left(x_{1},\,\boldsymbol{p}_{T}-\boldsymbol{k}_{T}\right)\int_{0}^{1}dz\int_{0}^{1}dz'\label{FD1-2}\\
 &  & \times\,\,\,\int\frac{d^{2}r_{1}}{4\pi}\,\int\frac{d^{2}r_{2}}{4\pi}e^{i\left(\boldsymbol{r}_{1}-\boldsymbol{r}_{2}\right)\cdot\boldsymbol{k}_{T}}\,\Psi_{\bar{Q}Q}^{\dagger}\left(r_{2},\,z,\,p_{T}\right)\Psi_{\bar{Q}Q}\left(r_{1},\,z,\,p_{T}\right)\nonumber \\
 &  & \times N_{M}^{({\rm SD})}\left(x_{2}(y);\,\vec{r}_{1},\,\vec{r}_{2}\right)+\left(x_{1}\leftrightarrow x_{2}\right),\nonumber \\
 &  & x_{1,2}\approx\frac{\sqrt{m_{M}^{2}+\langle p_{\perp M}^{2}\rangle}}{\sqrt{s}}e^{\pm y}
\end{eqnarray}
where $y$ and $\boldsymbol{p}_{T}$ are the rapidity and transverse
momenta of the produced heavy quark, in the center-of-mass frame of
the colliding protons; $\boldsymbol{k}_{T}$ is the transverse momentum
of the heavy quark; $g\left(x_{1},\,\boldsymbol{p}_{T}\right)$ in
the first line of~(\ref{FD1-2}) is the unintegrated gluon PDF; $\Psi_{g\to\bar{Q}Q}(r,\,z)$
is the light-cone wave function of the $\bar{Q}Q$ pair with transverse
separation between quarks $r$ and the light-cone fraction of the
momentum carried by the quark $z$. For $\Psi_{g\to\bar{Q}Q}(r,\,z)$
we use standard perturbative expressions~\cite{Rezaeian:2012ji}
\begin{align}
\Psi_{T}^{\dagger}\left(r_{2},\,z,\,Q^{2}\right)\Psi_{T}\left(r_{1},\,z,\,Q^{2}\right) & =\frac{\alpha_{s}N_{c}}{2\pi^{2}}\left\{ \epsilon_{f}^{2}\,K_{1}\left(\epsilon_{f}r_{1}\right)K_{1}\left(\epsilon_{f}r_{2}\right)\left[e^{i\theta_{12}}\,z^{2}+e^{-i\theta_{12}}(1-z)^{2}\right]\right.\\
 & \left.+m_{f}^{2}K_{0}\left(\epsilon_{f}r_{1}\right)K_{0}\left(\epsilon_{f}r_{2}\right)\right\} ,\nonumber \\
\Psi_{L}^{\dagger}\left(r_{2},\,z,\,Q^{2}\right)\Psi_{L}\left(r_{1},\,z,\,Q^{2}\right) & =\frac{\alpha_{s}N_{c}}{2\pi^{2}}\,\left\{ 4Q^{2}z^{2}(1-z)^{2}K_{0}\left(\epsilon_{f}r_{1}\right)K_{0}\left(\epsilon_{f}r_{2}\right)\right\} ,
\end{align}
\begin{equation}
\epsilon_{f}^{2}=z\,(1-z)\,Q^{2}+m_{f}^{2}
\end{equation}
\begin{equation}
\left|\Psi^{(f)}\left(r,\,z,\,Q^{2}\right)\right|^{2}=\left|\Psi_{T}^{(f)}\left(r,\,z,\,Q^{2}\right)\right|^{2}+\left|\Psi_{L}^{(f)}\left(r,\,z,\,Q^{2}\right)\right|^{2}
\end{equation}
The meson production amplitude $N_{M}$ depends on the mechanism of
the $Q\bar{Q}$ pair formation. For the case of the single-diffractive
production, as we demonstrate in the Appendix~\ref{sec:Derivation},
the contribution to the cross-section is given by

\begin{align}
N_{M}^{({\rm SD})}\left(x,\,z,\,\vec{\boldsymbol{r}}_{1},\,\vec{\boldsymbol{r}}_{2}\right)\approx\int d^{2}\boldsymbol{b} & \left[\mathcal{N}_{+}\left(x,\,z,\,\boldsymbol{r}_{1},\,\boldsymbol{b}\right)\left(\frac{N_{c}}{4}\right)+\mathcal{N}\left(x,\,\boldsymbol{r}_{1},\,\boldsymbol{b}\right)\left(\frac{N_{c}^{2}-4}{4N_{c}}+\frac{1}{6}\right)\right]\times\label{eq:N3Direct}\\
\times & \left[\mathcal{N}_{+}\left(x,\,z,\,\boldsymbol{r}_{2},\,\boldsymbol{b}\right)\left(\frac{N_{c}}{4}\right)+\mathcal{N}\left(x,\,\boldsymbol{r}_{2},\,\boldsymbol{b}\right)\left(\frac{N_{c}^{2}-4}{4N_{c}}+\frac{1}{6}\right)\right].\nonumber 
\end{align}

where 
\begin{align}
\mathcal{N}_{+}\left(x,\,z,\,\boldsymbol{r},\,\boldsymbol{b}\right) & \equiv2\mathcal{N}\left(x,\,z\boldsymbol{r},\,\boldsymbol{b}\right)+2\mathcal{N}\left(x,\,\bar{z}\boldsymbol{r},\,\boldsymbol{b}\right)-\frac{1}{2}\mathcal{N}\left(x,\,\boldsymbol{r},\,\boldsymbol{b}\right),
\end{align}
and $\mathcal{N}\left(x,\,\boldsymbol{r},\,\boldsymbol{b}\right)$
is the color singlet dipole cross-section with explicit dependence
on impact parameter $\boldsymbol{b}$.

In the heavy quark mass limit the main contribution to the integrals
in~(\ref{FD1-2}) comes from small dipoles of size $r\lesssim m_{Q}^{-1}$.
In widely used phenomenological dipole parametrizations~\cite{Kowalski:2003hm,Kowalski:2006hc,Rezaeian:2012ji,RESH}
it is expected that the $b$- and $r$-dependence factorize in this
limit, 
\begin{equation}
\mathcal{N}\left(x,\,\boldsymbol{r},\,\boldsymbol{b}\right)\approx N\left(x,\,\boldsymbol{r}\right)T(\boldsymbol{b}),\label{eq:fact}
\end{equation}
where the transverse profile $T(b)$ is normalized as $\int d^{2}b\,T(b)=1$,
and $N\left(x,\,\boldsymbol{r}\right)$ is the dipole cross-section
integrated over impact parameter. In this approximation we may rewrite~(\ref{eq:N3Direct})
as 
\begin{align}
N_{M}^{({\rm SD})}\left(x,\,z,\,\vec{\boldsymbol{r}}_{1},\,\vec{\boldsymbol{r}}_{2}\right)\approx & \kappa\left[N_{+}\left(x,\,z,\,\boldsymbol{r}_{1}\right)\left(\frac{N_{c}}{4}\right)+N\left(x,\,\boldsymbol{r}_{1}\right)\left(\frac{N_{c}^{2}-4}{4N_{c}}+\frac{1}{6}\right)\right]\times\label{eq:N3Direct-1}\\
\times & \left[N_{+}\left(x,\,z,\,\boldsymbol{r}_{2}\right)\left(\frac{N_{c}}{4}\right)+N\left(x,\,\boldsymbol{r}_{2}\right)\left(\frac{N_{c}^{2}-4}{4N_{c}}+\frac{1}{6}\right)\right],\nonumber 
\end{align}

where 
\begin{align}
N_{+}\left(x,\,z,\,\boldsymbol{r}\right) & \equiv\int d^{2}b\,\mathcal{N}_{+}\left(x,\,z,\,\boldsymbol{r},\,\boldsymbol{b}\right)=2N\left(x,\,z\boldsymbol{r}\right)+2N\left(x,\,\bar{z}\boldsymbol{r}_{1}\right)-\frac{1}{2}N\left(x,\,\boldsymbol{r}\right),\\
\kappa & =\int d^{2}\boldsymbol{b}\,T^{2}(b).
\end{align}
As could be seen from the structure of~(\ref{eq:N3Direct}), it is
a higher twist ($\sim\mathcal{O}\left(r^{2}\right)$ ) contribution
compared to the amplitude of inclusive production, and thus should
have stronger suppression at large $p_{T}$.

The $p_{T}$-integrated cross-section gets contributions only from
dipoles with $\vec{\boldsymbol{r}}_{1}=\vec{\boldsymbol{r}}_{2}=\vec{\boldsymbol{r}}$
in the integrand. For this case it is possible to show that the gluon
uPDF $x_{1}\,g\left(x_{1},\,\boldsymbol{p}_{T}-\boldsymbol{k}_{T}\right)$
is replaced with the integrated gluon PDF $x_{g}G\left(x_{g},\mu_{F}\right)$
taken at the scale $\mu_{F}\,\approx2\,m_{Q}$. In the LHC kinematics
at central rapidities this scale significantly exceeds the saturation
scale $Q_{s}(x)$, which justifies the dominance of the three-pomeron
approximation. However, in the small-$x$ kinematics there are sizeable
nonlinear corrections to the evolution in the dipole approach. In
this kinematics the corresponding scale $\mu_{F}$ should be taken
at the saturation momentum $Q_{s}$. The gluon PDF $x_{1}G\left(x_{1},\,\mu_{F}\right)$
in this approach is closely related to the dipole scattering amplitude
$N\left(x,\,\boldsymbol{r}\right)=\int d^{2}b\,N\left(x,\,\boldsymbol{r},\,\boldsymbol{b}\right)$
as~\cite{KOLEB,THOR} 
\begin{equation}
\frac{C_{F}}{2\pi^{2}\bar{\alpha}_{S}}N\left(x,\,\boldsymbol{r}\right)=\int\frac{d^{2}k_{T}}{k_{T}^{4}}\phi\left(x,\,\boldsymbol{k}_{T}\right)\,\Bigg(1-e^{i\boldsymbol{k}_{T}\cdot\boldsymbol{r}}\Bigg);~~~~x\,G\left(x,\,\mu_{F}\right)=\int_{0}^{\mu_{F}}\frac{d^{2}k_{T}}{k_{T}^{2}}\phi\left(x,\,\boldsymbol{k}_{T}\right),\label{GN1-1}
\end{equation}
Eq. (\ref{GN1-1}) can be inverted and gives the gluon uPDF in terms
of the dipole amplitude, 
\begin{equation}
xG\left(x,\,\mu_{F}\right)\,\,=\,\,\frac{C_{F}\mu_{F}}{2\pi^{2}\bar{\alpha}_{S}}\int d^{2}r\,\frac{J_{1}\left(r\,\mu_{F}\right)}{r}\nabla_{r}^{2}N\left(x,\,\boldsymbol{r}\right).\label{GN2-1}
\end{equation}

The corresponding unintegrated gluon PDF can be rewritten as~\cite{Kimber:2001sc}
\begin{equation}
x\,g\left(x,\,k^{2}\right)=\left.\frac{\partial\,}{\partial\mu_{F}^{2}}xG\left(x,\,\mu_{F}\right)\right|_{\mu_{F}^{2}=k^{2}},\label{eq:GN3-1}
\end{equation}
which allows to express the single diffractive cross-section in terms
of only the dipole amplitude. The expression~(\ref{eq:GN3-1}) will
be used below in Section~\ref{sec:MultiplicityGeneralities} for
extension of our results to high-multiplicity events.

\subsection{Gap survival factors}

\label{subsec:LRGSF}The rapidity gap between the recoil proton and
the produced heavy meson might be filled potentially by products of
various secondary processes, as shown schematically by the colored
vertical and inclined ovals in Figure~\ref{fig:SD}. As was demonstrated
in~\cite{Martin:2008nx,Khoze:2018kna,Ryskin:2009tj,Khoze:2000},
the effect of these factors is significant at high energies and might
decrease the observed yields (\emph{i.e}. probability of non-observation
of particles in the gap) by more than an order of magnitude\textbf{\textcolor{red}{~}}\cite{Khoze:2000,Khoze:2017sdd}.
This suppression is due to soft interactions between the colliding
protons and thus is not related to the particles produced due to hard
interactions. The evaluation of this suppression conventionally follows
the ideas of Good-Walker~\cite{GoodWalker:1960}, which are usually
implemented in the context of different models~(see for review~\cite{Gotsman:2005wa,Gotsman:2005rt,Ryskin:2009qf,Bialas:1995bs}).
Technically, all these approaches perform evaluations in eikonal approximation,
and predict that the observables, which include large rapidity gaps,
are suppressed by a so-called gap survival factor, 
\begin{equation}
\left\langle S^{2}\right\rangle =\frac{\int d^{2}b\,\left|\mathcal{M}\left(b,\,s,\,...\right)\right|^{2}\exp\left(-\hat{\Omega}(b,\,s)\right)}{\int d^{2}b\,\left|\mathcal{M}\left(b,\,s,\,...\right)\right|^{2}},\label{eq:S2}
\end{equation}
where $\mathcal{M}(b,\,s,\,...)$ is the amplitude of the hard process,
$b$ is the impact parameter, and $\Omega$ is the opacity or optical
density. In a single-channel eikonal model the opacity $\Omega$ is
directly related to the cross-sections of total, elastic and inelastic
processes~\cite{Gotsman:2005rt}. It is expected that the energy
dependence of the function $\Omega$ is controlled by the Pomeron
intercept, $\Omega\sim s^{\alpha_{IP}-1}$, so the factor (\ref{eq:S2})
decreases as a function of energy. The single-channel model is very
simple, yet its predictions are at tension with experimental data~\cite{Khoze:2017sdd}.
More accurate description of data is achieved in multichannel extensions
of these models, which assume that after interaction with a soft Pomeron
the proton might convert into additional $N_{D}-1$ diffractive states.
In this basis, the soft pomeron interaction amplitude $\hat{\Omega}$
should be considered as an $N_{D}\times N_{D}$ matrix. As was discussed
in~\cite{Gotsman:2005wa,Gotsman:2005rt,Ryskin:2009qf}, for a good
description it is sufficient to choose $N_{D}=2$, with the common
parametrization for the matrix $\Omega_{ik}$ given in~\cite{Gotsman:1999ri}
and briefly summarized for the sake of completeness in Appendix~\ref{sec:SoftPomeron}.
For the single diffractive scattering the exponent in the expression~(\ref{eq:S2})
should be understood as a matrix element between $|pp\rangle$ and
$|p\,X\rangle$ states~\cite{Khoze:2014aca,Khoze:2000wk}. If $\Phi_{1}$
and $\Phi_{2}$ are eigenvalues of $\Omega_{ik}$ with eigenvalues
$\Omega_{1}$ and $\Omega_{2}$, then the matrix $\exp\left(-\hat{\Omega}(b,\,s)\right)$
reduces in this basis to a linear combination of factors $\sim e^{-\Omega_{a}(s,b)}$,
in which the coefficients can be fixed by projecting the proton and
diffractive states onto the eigenstates $\Phi_{1},\,\Phi_{2}$ of
the scattering matrix. For the single diffractive production the algorithm
for evaluation of the survival factor was introduced earlier for the
$pp\to pX$ process in~\cite{Khoze:2000wk}, yielding 
\begin{equation}
\exp\left(-\hat{\Omega}(b,\,s)\right)\to\mathcal{S}^{2}\left(s_{pp},\,b\right)\equiv\frac{1}{4(1+\lambda^{2})}\left((1+\lambda)^{3}e^{-(1+\lambda)^{2}\Omega}+(1-\lambda)^{3}e^{-(1-\lambda)^{2}\Omega}+2\left(1-\lambda^{2}\right)e^{-\left(1-\lambda^{2}\right)\Omega}\right),\label{eq:S2Int}
\end{equation}
where parameter $\Omega$ is related to eigenvalues $\Omega_{1,2}$
of the matrix $\Omega_{ik}$ as 
\begin{align}
\Omega & =\frac{\Omega_{1}+\Omega_{2}}{2},
\end{align}
and the parameter $\lambda$ stands for the ratio of the production
amplitude of diffractive state $X$ to the amplitude of elastic proton
scattering of the incident proton on a pomeron (see Appendix~\ref{sec:SoftPomeron}
for more details). In this paper we are interested only in events
without charged particles, produced at pseudorapidity $\eta<y$ (rapidity
gap between the recoil proton and heavy quarks), whereas the evaluation
of the survival factor in~(\ref{eq:S2},\ref{eq:S2Int}\ref{eq:S2Ave})
was performed under the assumption that there are no co-produced particles
in the whole rapidity range $\eta\in\left(-y_{{\rm max}},\,y_{{\rm max}}\right)$,
which is much stricter than needed in this problem. For this reason
we need to correct the estimate~(\ref{eq:S2Int}), using probabilistic
considerations. In what follows we'll use notations $P_{A}$ and $P_{B}$
for the probabilities to emit at least one charged particle in the
intervals $\eta<y$ and $\eta>y$ due to soft interaction of the colliding
protons; while $\bar{P}_{A}\equiv1-P_{A}$ and $\bar{P}_{B}\equiv1-P_{B}$
are the probabilities not to emit any particles in these intervals
(the gap survival factors on these intervals). We will also use the
notation $\bar{P}_{A\cup B}$ for the probability not to produce particles
in any of the intervals. The relation between the probabilities $\bar{P}_{A\cup B}$
and $\bar{P}_{A},\,\bar{P}_{B}$ depends crucially on possible correlations
between particles from different rapidity intervals. Such correlations
have been studied in the literature~\cite{Alver:2007wy,Eggert:1974ek,Khachatryan:2010gv},
and it is known that they are small when the separation between the
bins is larger than 1-2 units in rapidity. If we neglect completely
such correlations, the probabilities are related as $\bar{P}_{A\cup B}=\bar{P}_{A}\bar{P}_{B}$,
which implies that the survival factor should scale with the length
of the rapidity bin as $S^{2}\left(\Delta\eta\right)\sim{\rm const}^{\Delta\eta}$.
For the single diffractive production of heavy mesons we require that
no particles are produced with $\eta<y$, although we do not impose
any conditions for $\eta>y$ (so we do not need to introduce the gap
survival factor in this region). This implies that the overall survival
factor~(\ref{eq:S2Int}) should be adjusted as 
\begin{equation}
\mathcal{S}^{2}\to S^{2}\left(s_{pp},\,b\right)=\left(\mathcal{S}^{2}\right)^{\frac{\Delta y}{2y_{{\rm max}}}}\gtrsim\mathcal{S}^{2},
\end{equation}
where $\Delta y$ is the width of the rapidity gap interval, and $y_{{\rm max}}=-\frac{1}{2}\ln\left(m_{Q,T}^{2}/s\right)$
is the largest possible rapidity of heavy quarks. This factor $S^{2}\left(s_{pp},\,b\right)$
should be included into the expressions~(\ref{FD1-2},\ref{eq:N3Direct})
from the previous Section~(\ref{subsec:3Pom-1}).

In the heavy quark mass limit the dipoles are small, $r\lesssim m_{Q}^{-1}$,
and we may use a factorized approximation~(\ref{eq:fact}). The convolution
of~$S^{2}\left(s_{pp},\,b\right)$ with impact parameter dependent
cross-section can be simplified in this limit and yields for the suppression
factor a much simpler expression 
\begin{equation}
\left\langle S^{2}\right\rangle \approx\frac{\int d^{2}b\,T^{2}(b)\,S^{2}\left(s_{pp},\,b\right)}{\int d^{2}b\,T^{2}(b)},\label{eq:S2Ave}
\end{equation}
which depends only on the energy (Mandelstam variable) $s_{pp}$ of
the collision, but does not depend on masses nor kinematics of the
produced heavy quarks.

\section{Numerical results}

\label{sec:Numer} 
\begin{figure}
\includegraphics[width=9cm]{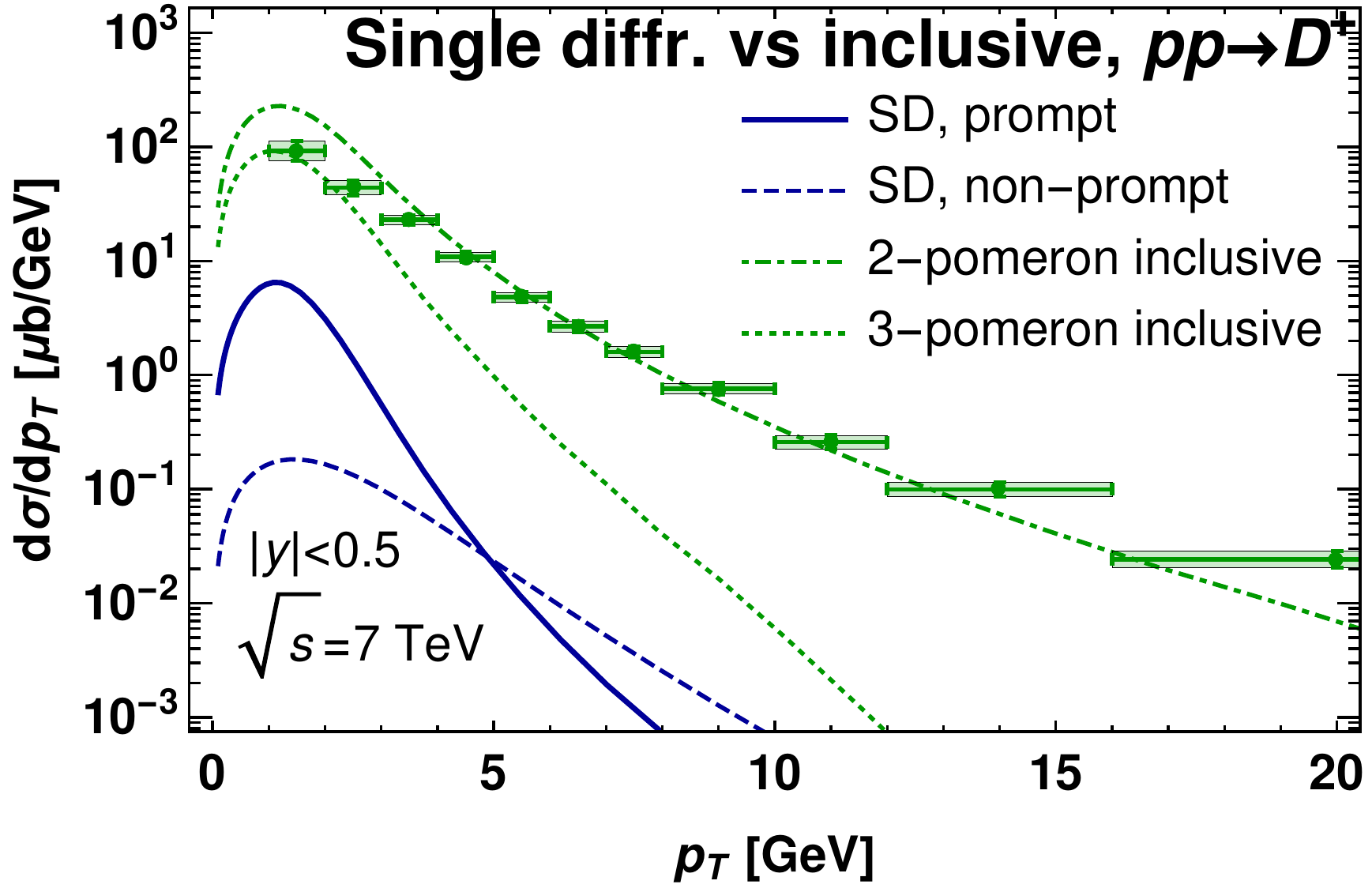}\includegraphics[width=9cm]{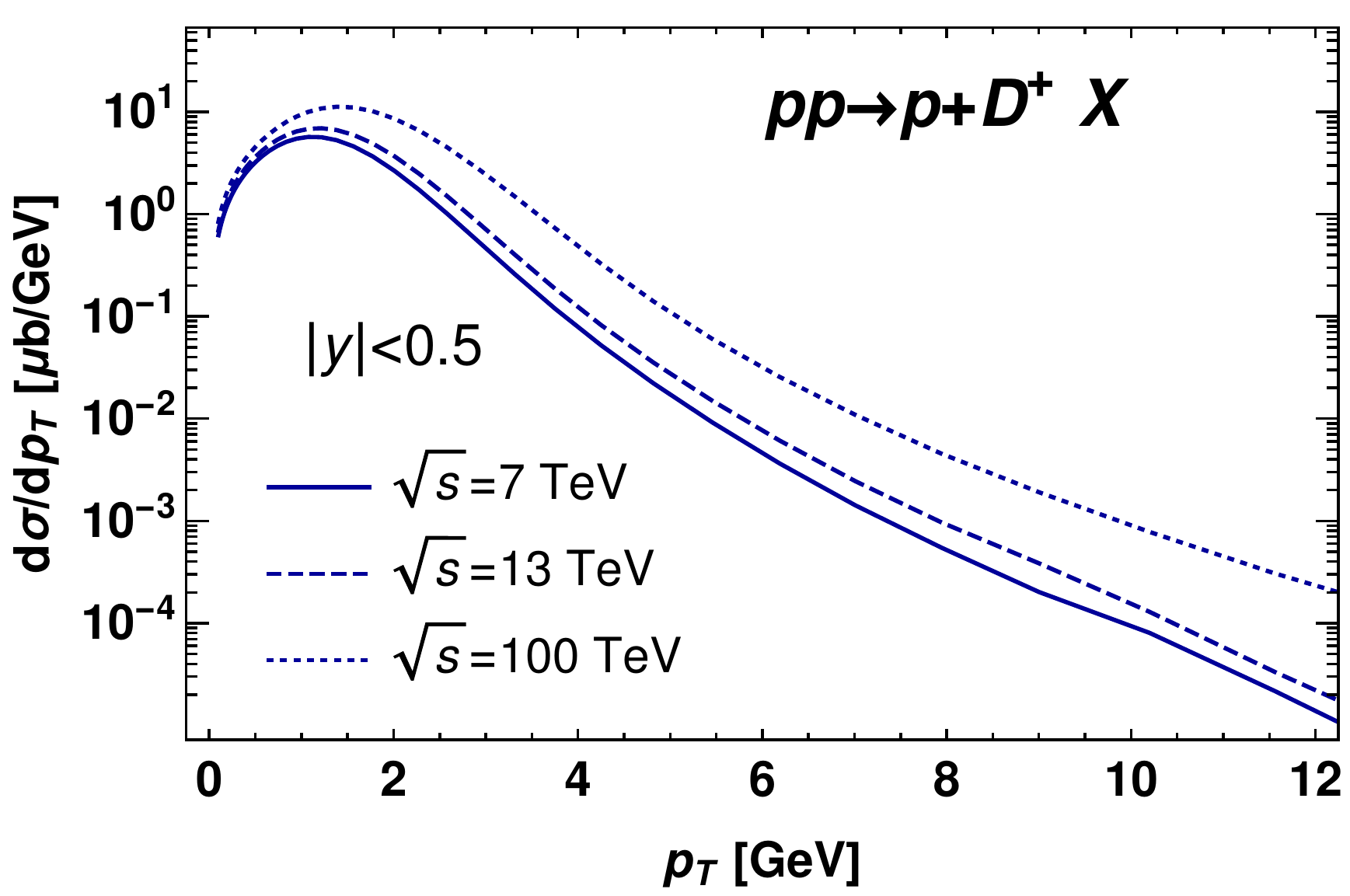}

\caption{\label{fig:pTDependence}The cross-section $d\sigma/dp_{T}$ of the
single diffractive production of $D^{+}$-mesons. Integration over
the rapidity bin $|y|<0.5$ is implied. Left plot: Comparison with
inclusive production in the LHC kinematics for $\sqrt{s}=7$~TeV
(theory and experiment). The curves with labels ``SD, prompt'' and
``SD, non-prompt'' correspond to single diffractive contributions
to $D$-meson yields from the fragmentation of the $c$ and $b$ quarks
respectively. The curves marked ``2-pomeron inclusive'' and ``3-pomeron
inclusive'' stand for the contributions of 2- and 3-pomeron fusion
mechanisms to inclusive $D$-meson yields respectively (see a short
overview in Appendix~\ref{subsec:InclusiveSummary} and more detailed
discussion in~\cite{Schmidt:2020fgn}). The experimental data are
for inclusive production from~\cite{Acharya:2017jgo}. Right plot:
$\sqrt{s}$-dependence of the data in the kinematics of LHC and the
planned Future Cicular Collider (FCC)~\cite{Mangano:2017tke}. For
other $D$-mesons the $p_{T}$-dependence has a very similar shape,
yet differs numerically by a factor of two.}
\label{Diags_DMesons} 
\end{figure}

\begin{figure}
\includegraphics[width=9cm]{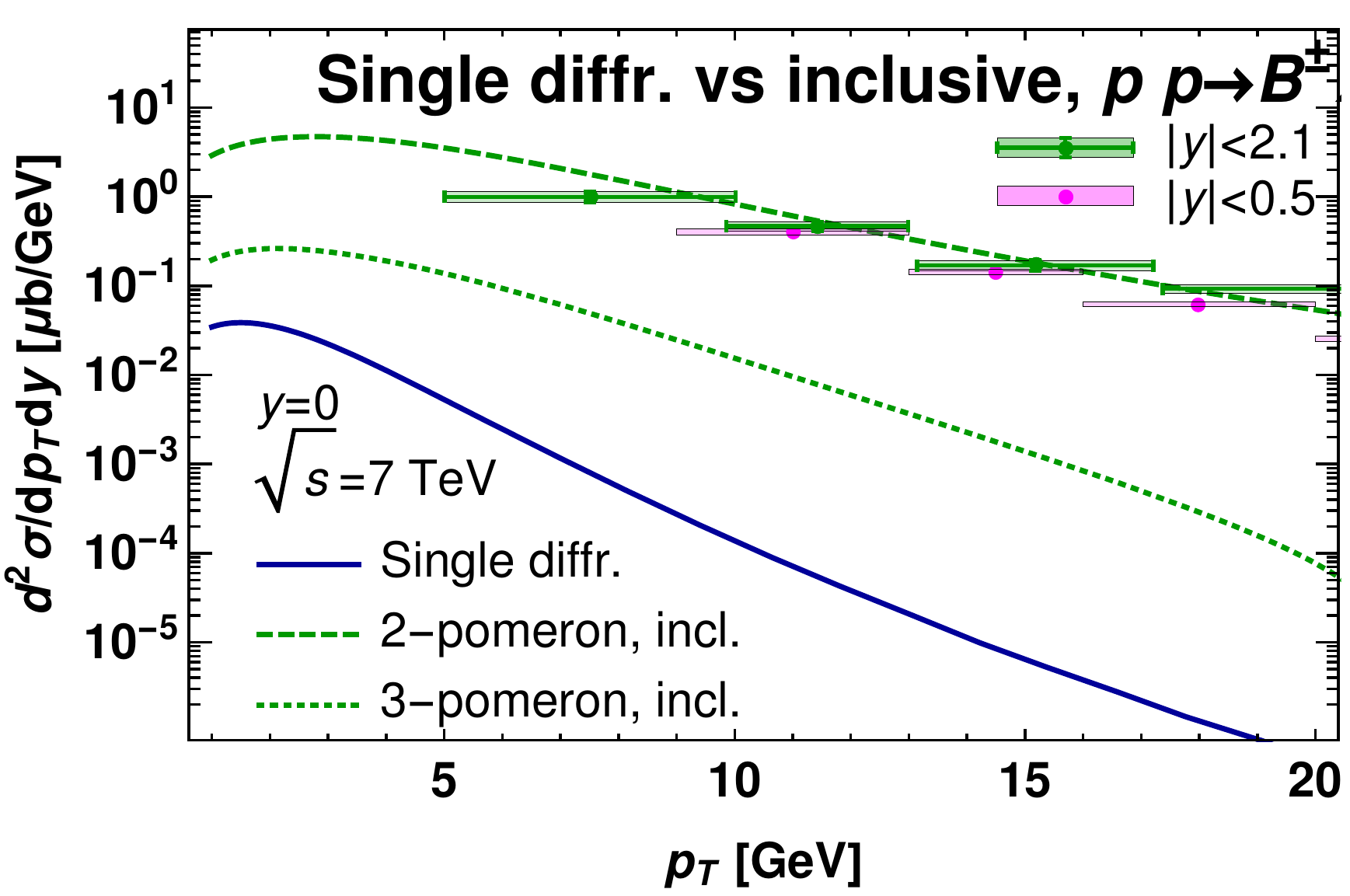}\includegraphics[width=9cm]{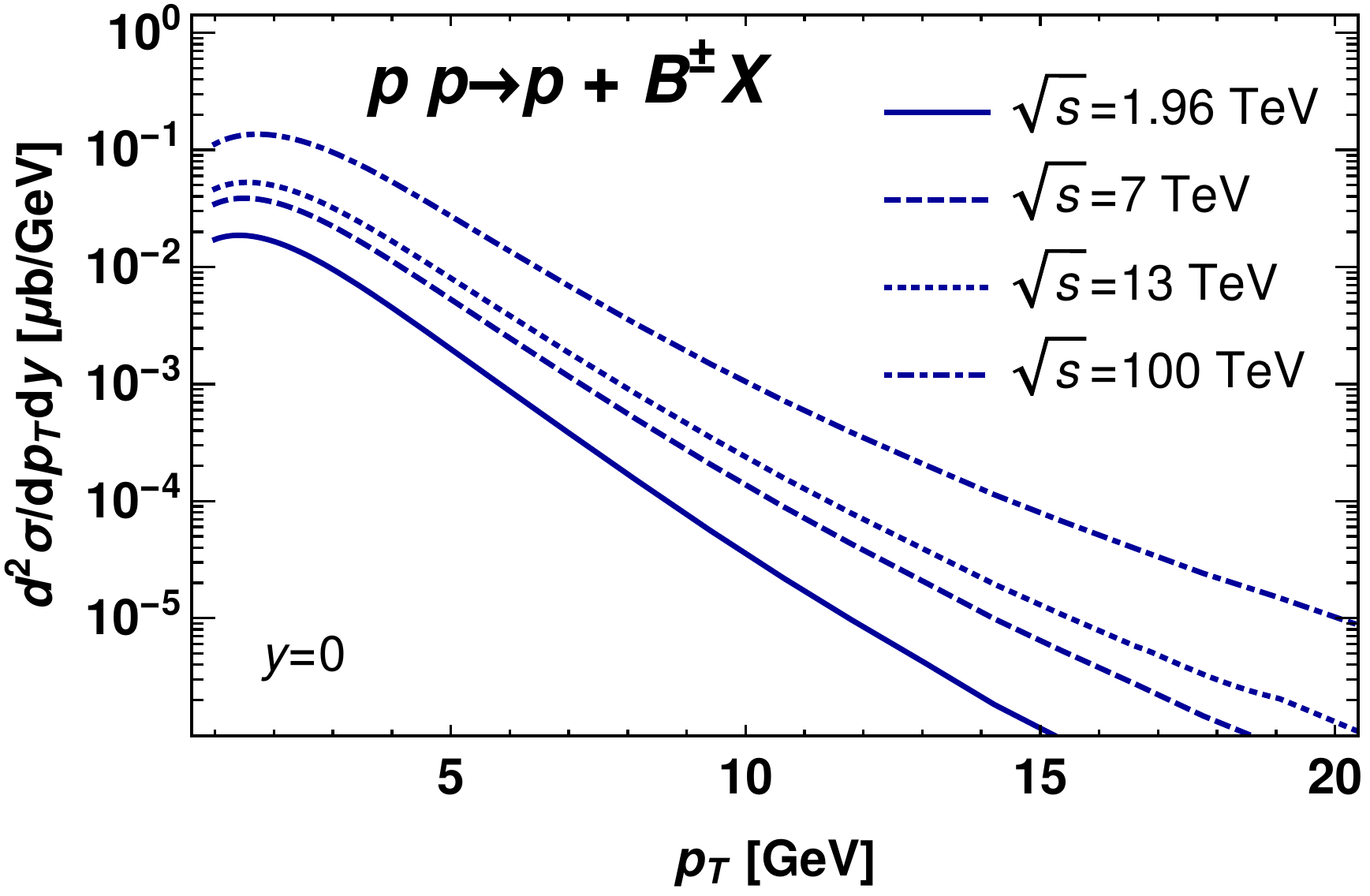}

\caption{\label{fig:pTDependence-1}Cross-section for the single diffractive
$B^{\pm}$-mesons production. Left plot: Comparison of single diffractive
predictions with inclusive cross-sections (experimental and theoretical
results). The theoretical curves marked ``2-pomeron incl.'' and
``3-pomeron incl.'' stand for the additive contributions from 2-
and 3-pomeron fusion mechanisms respectively (see~\cite{Schmidt:2020fgn}
and a short discussion in Appendix~\ref{subsec:InclusiveSummary}).The
experimental data are for inclusive production from CMS~\cite{Khachatryan:2016csy}(``$|y|<2.1$``
data points) and ATLAS~\cite{ATLAS:2013cia}(``$|y|<0.5$'' data
points). For some experimentally measured results bin-integrated cross-sections
$d\sigma/dp_{T}$ was converted into $d\sigma/dp_{T}dy$ dividing
by the width of the rapidity bin (this is justified since in LHC kinematics
at central rapidities $y\approx0$ the cross-section is flat). Right
plot: The $p_{T}$-dependence of the cross-section $d\sigma/dy\,dp_{T}$
for several energies $\sqrt{s}$. }
\label{Diags_BMesons} 
\end{figure}

\begin{figure}
\includegraphics[width=9cm]{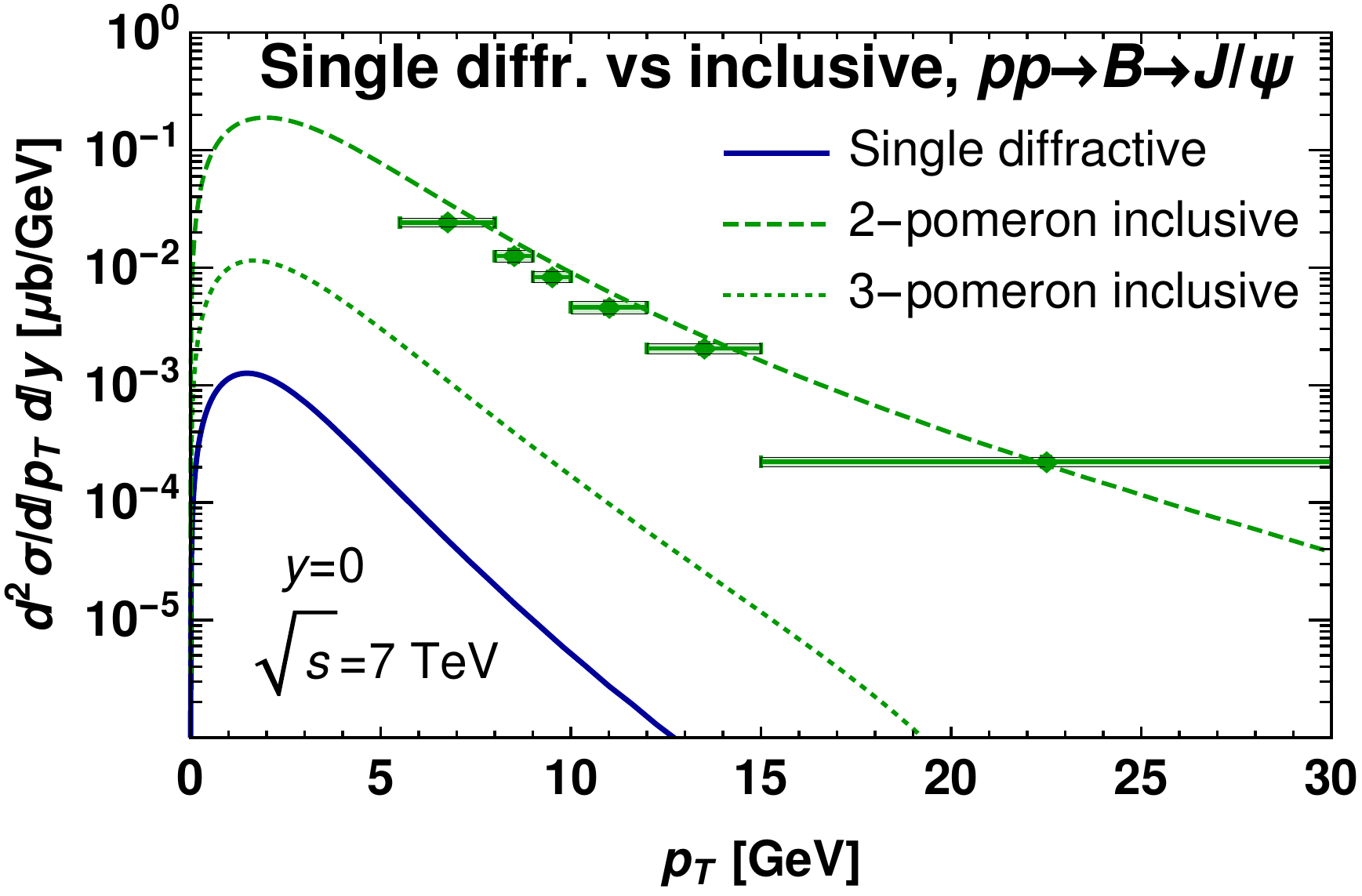}\includegraphics[width=9cm]{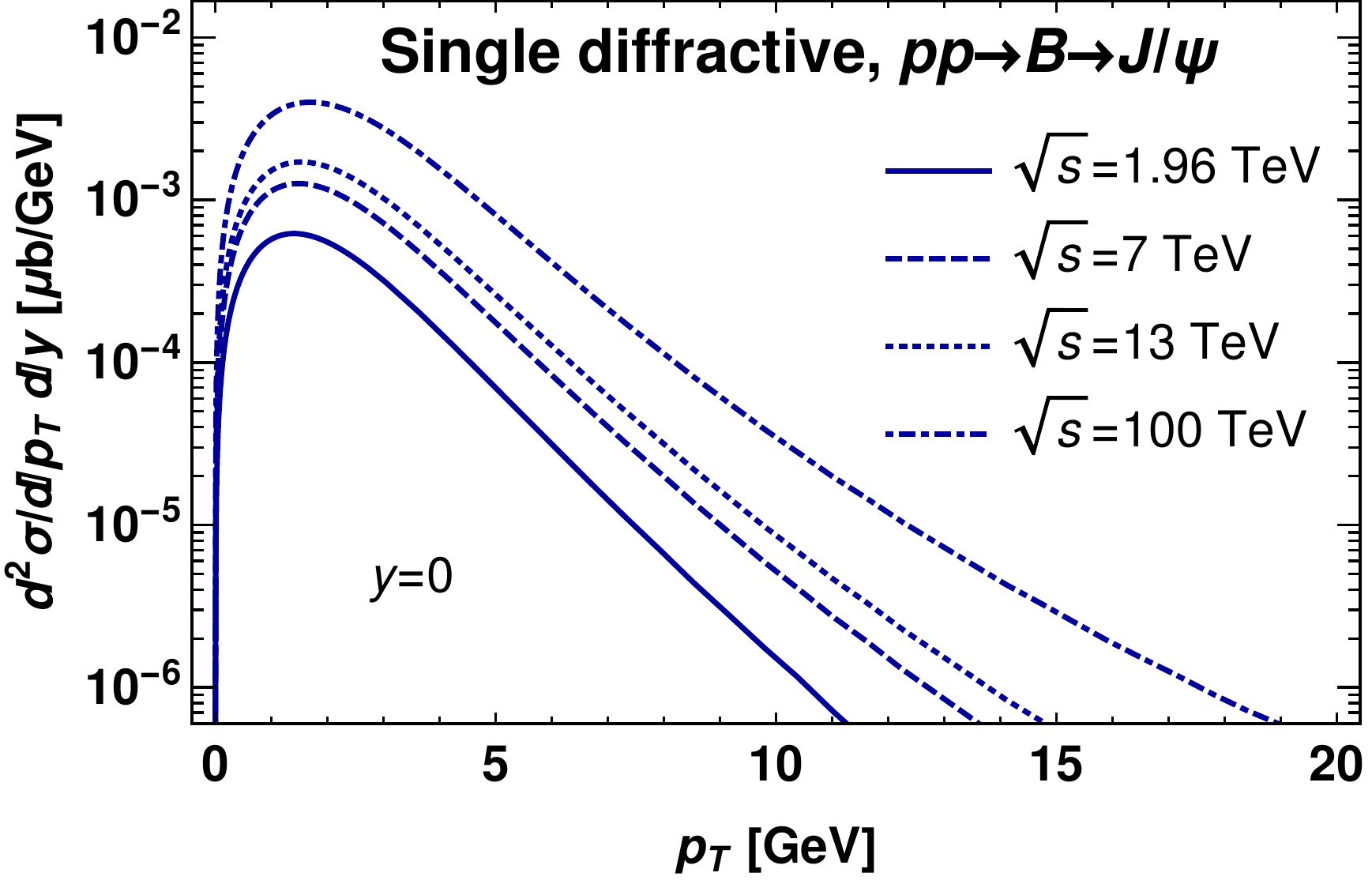}

\caption{\label{fig:pTDependence-nonprompt}Cross-section for the single diffractive
non-prompt $J/\psi$-mesons production. Left plot: Comparison of single
diffractive predictions with inclusive cross-sections (experimental
and theoretical results). The theoretical curves marked ``2-pomeron
inclusive'' and ``3-pomeron inclusive'' stand for the additive
contributions from 2- and 3-pomeron fusion mechanisms respectively
(see~\cite{Schmidt:2020fgn} and a short discussion in Appendix~\ref{subsec:InclusiveSummary}).The
experimental data are for inclusive production from CMS~\cite{Chatrchyan:2011k}.
Right plot: The $p_{T}$-dependence of the cross-section $d\sigma/dy\,dp_{T}$
for several energies $\sqrt{s}$. }
\label{Diags_JMesons-2} 
\end{figure}

For our numerical evaluations here and in what
follows we will we use the impact parameter ($b$) dependent ``bCGC''
parametrization of the dipole cross-section~\cite{Kowalski:2006hc,RESH}
\begin{align}
N\left(x,\,\boldsymbol{r},\,\boldsymbol{b}\right) & =\left\{ \begin{array}{cc}
N_{0}\,\left(\frac{r\,Q_{s}(x)}{2}\right)^{2\gamma_{{\rm eff}}(r)}, & r\,\le\frac{2}{Q_{s}(x)}\\
1-\exp\left(-\mathcal{A}\,\ln\left(\mathcal{B}r\,Q_{s}\right)\right), & r\,>\frac{2}{Q_{s}(x)}
\end{array}\right.~,\label{eq:CGCDipoleParametrization}\\
 & \mathcal{A}=-\frac{N_{0}^{2}\gamma_{s}^{2}}{\left(1-N_{0}\right)^{2}\ln\left(1-N_{0}\right)},\quad\mathcal{B}=\frac{1}{2}\left(1-N_{0}\right)^{-\frac{1-N_{0}}{N_{0}\gamma_{s}}},\\
 & Q_{s}(x,\,\boldsymbol{b})=\left(\frac{x_{0}}{x}\right)^{\lambda/2}T_{G}(b),\,\,\gamma_{{\rm eff}}(r)=\gamma_{s}+\frac{1}{\kappa\lambda Y}\ln\left(\frac{2}{r\,Q_{s}(x)}\right),\label{eq:gamma_eff}\\
\gamma_{s} & =0.66,\quad\lambda=0.206,\quad x_{0}=1.05\times10^{-3},\quad T_{G}(b)=\exp\left(-\frac{b^{2}}{2\gamma_{s}B_{{\rm CGC}}}\right).
\end{align}
In Figures~\ref{fig:pTDependence},~\ref{fig:pTDependence-1} and
\ref{fig:pTDependence-nonprompt} we show the production cross-sections
of the $D$-mesons, $B$-mesons and non-prompt $J/\psi$ mesons. We
can see that in the small-$p_{T}$ region, which encompasses most
of the events, the single diffraction production constitutes approximately
one per cent of the inclusive cross-section. In the large-$p_{T}$
region the contribution from the single diffractive production is
strongly suppressed since it is formally a higher twist effect.

To the best of our knowledge there is no direct experimental data
for the cross-sections of the suggested process. The diffractive production
of $B$-mesons has been studied earlier by the CDF collaboration in~\cite{Affolder:1999hm},
although the results are only available for the ratio of the integrated
cross-sections of diffractive and inclusive processes, 
\begin{equation}
R_{\bar{b}b}^{({\rm diff.})}(s)\equiv\frac{\sigma_{B^{+}}^{{\rm diff}}(s)}{\sigma_{B^{+}}^{{\rm incl}}(s)}.\label{eq:RDiff-1}
\end{equation}
For energy $\sqrt{s}=1.8\,{\rm TeV}$ it was found that 
\begin{equation}
R_{\bar{b}b}^{({\rm diff.})}(\sqrt{s}=1.8\,{\rm TeV})=\left(0.62\pm0.19\pm0.16\right)\%.\label{eq:RDiff}
\end{equation}
In the Table~\ref{tab:Ratios} we present our theoretical expectations
for this value. For Tevatron kinematics the model prediction $R_{\bar{b}b}^{({\rm diff.})}\approx0.4\,\%$
agrees with~(\ref{eq:RDiff}), within uncertainty of experimental
data~(\ref{eq:RDiff}). As we can see from the same Table~\ref{tab:Ratios},
in LHC kinematics the ratio~(\ref{eq:RDiff-1}) is approximately
of the same order. The smallness of the values in the Table~\ref{tab:Ratios}
is due to the fact that the production of heavy quark in single diffraction
events is formally a higher twist effect, and thus has an additional
suppression by the factor $\sim\left(\Lambda_{{\rm QCD}}/m_{Q}\right)^{2}$.
While the absolute cross-sections of single diffractive and inclusive
production increase as a function of energy, the ratio~(\ref{eq:RDiff})
slowly \emph{decreases} due to energy dependence of the gap survival
factor in single-diffractive cross-section.

\begin{table}
\begin{tabular}{|c|c|c|c|}
\hline 
$\sqrt{s}$  & $R_{\bar{c}c}^{({\rm diff})}$  & $R_{\bar{b}b}^{({\rm diff})}$  & $R_{J/\psi}^{({\rm diff})}$\tabularnewline
\hline 
1.8~TeV  & 2.20~\%  & 0.40~\%  & 0.57\%\tabularnewline
\hline 
7~TeV  & 1.87~\%  & 0.33~\%  & 0.45\%\tabularnewline
\hline 
13~TeV  & 1.59~\%  & 0.30~\%  & 0.40\%\tabularnewline
\hline 
\end{tabular}\caption{\label{tab:Ratios}Values of the ratio of single diffractive and inclusive
productions cross-sections, as defined in~(\ref{eq:RDiff-1}), in
Tevatron and LHC kinematics. The second and the third columns correspond
to the $c$- and $b$-quarks ($R_{\bar{c}c}^{({\rm diff})}$ and $R_{\bar{b}b}^{({\rm diff.})}$
respectively). The last column $R_{J/\psi}^{({\rm diff})}$ is for
the non-prompt $J/\psi$ production.}
\end{table}

\begin{figure}
\includegraphics[width=9cm]{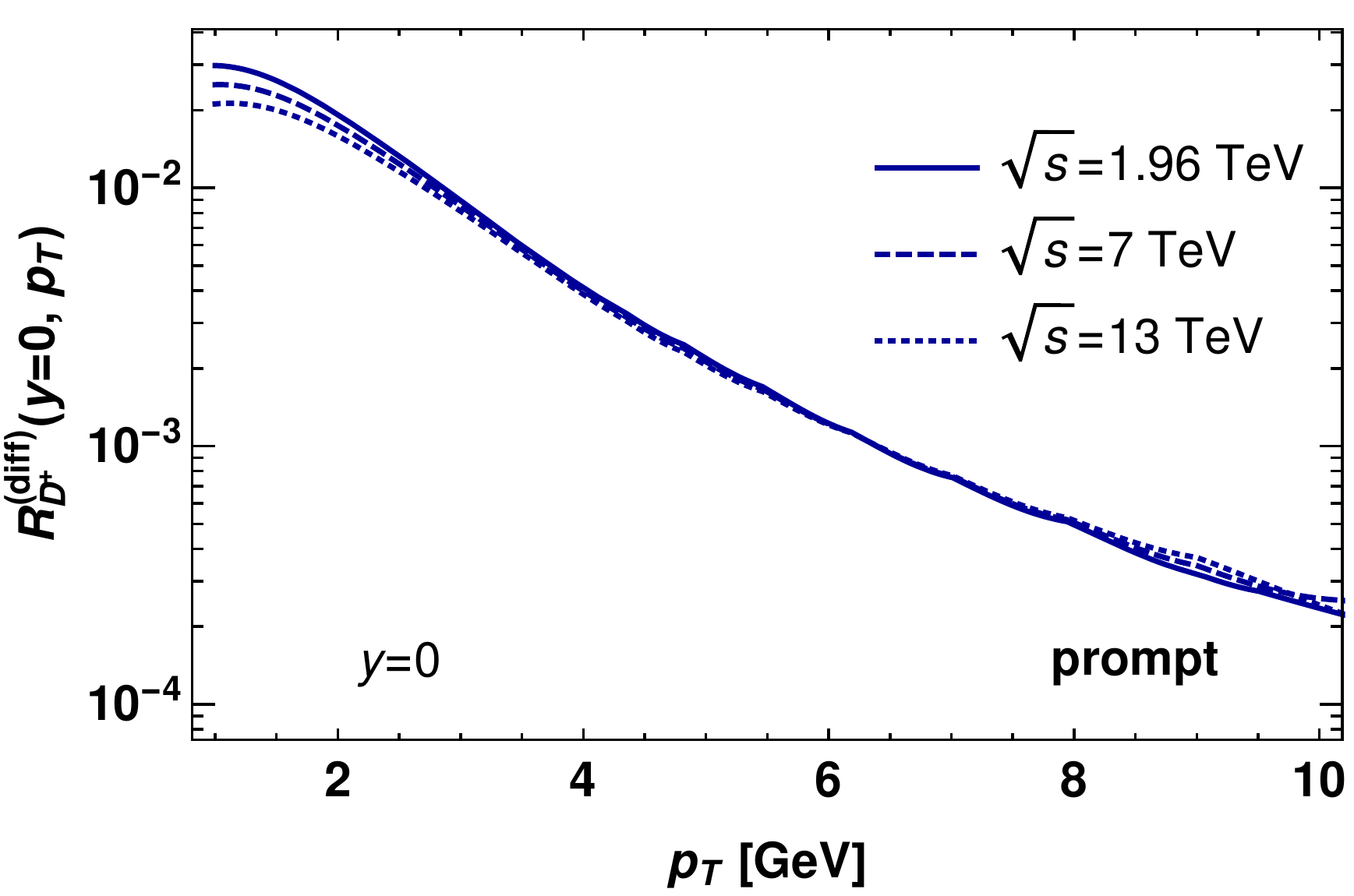}\includegraphics[width=9cm]{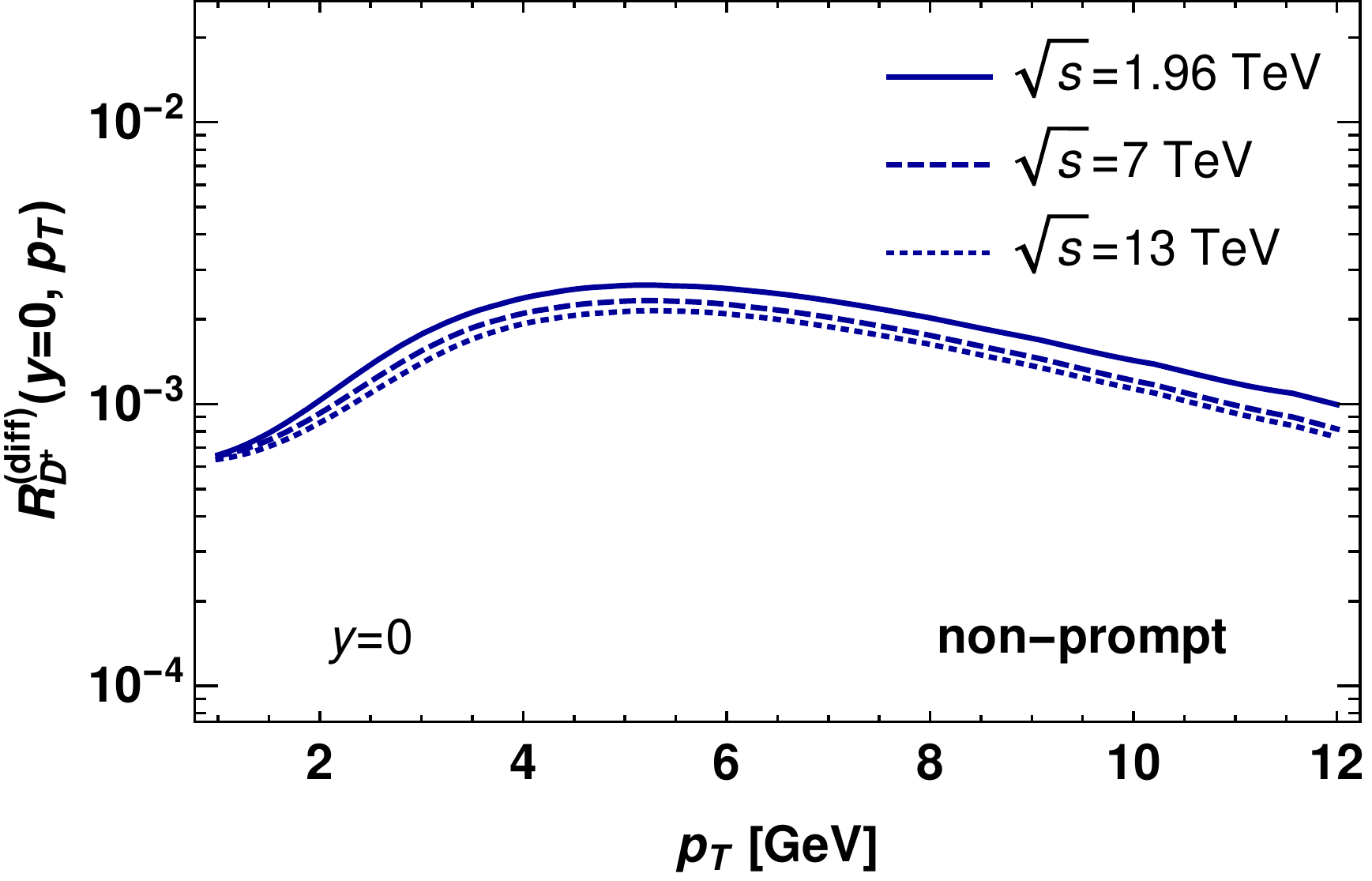}

\caption{\label{fig:pTDependenceRatio}The ratio of single diffractive to inclusive
production cross-sections, as defined in~(\ref{eq:RDiff-2}). The
left plot corresponds to the prompt production (from $c\to D$ fragmentation),
and the right plot is for the non-prompt mechanism (from $b\to D$
fragmentation). For the sake of definiteness we considered $D^{+}$
mesons; for other $D$-mesons the results have a very similar shape. }
\label{Diags_BMesons-1} 
\end{figure}

\begin{figure}
\includegraphics[width=9cm]{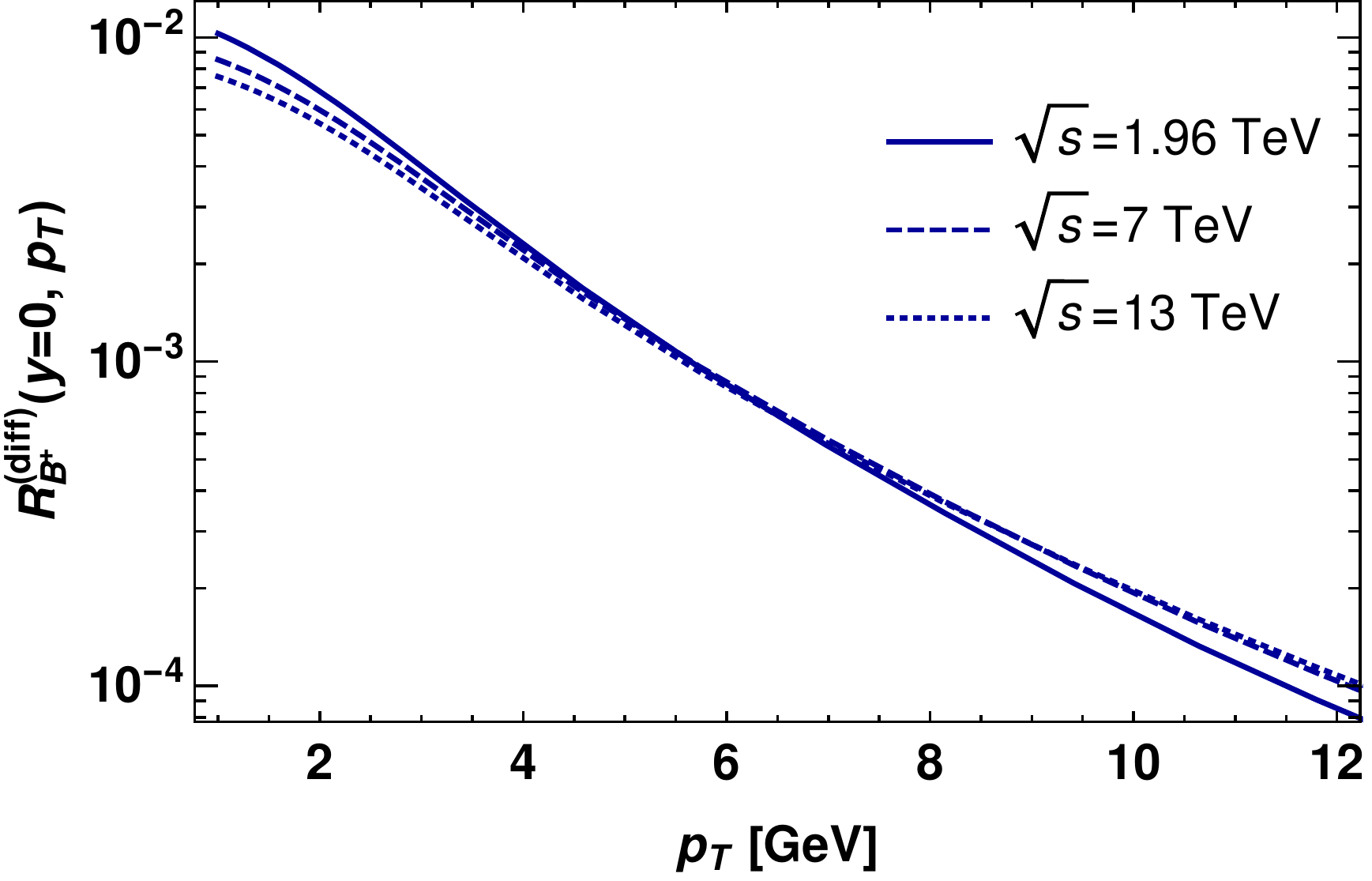}\includegraphics[width=9cm]{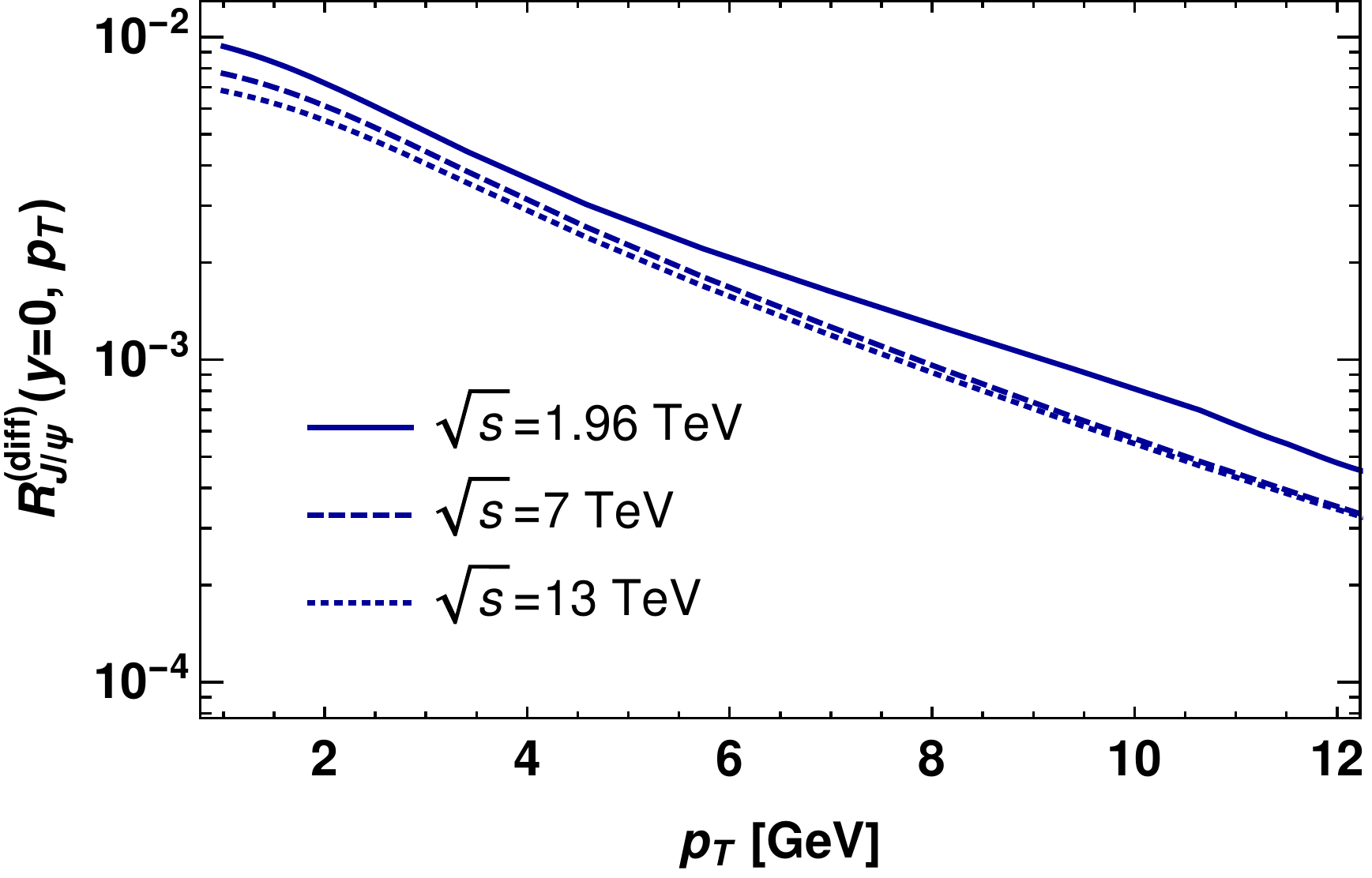}

\caption{\label{fig:pTDependenceRatio-2}The ratio of single diffractive to
inclusive production cross-sections, as defined in~(\ref{eq:RDiff-2}).
The left plot is for the $B$ mesons, the right panel is for non-prompt
production of $J/\psi$-mesons.}
\label{Diags_BMesons-1-1} 
\end{figure}

We extended the definition~(\ref{eq:RDiff-1}) and analyzed the ratio
of differential cross-sections, 
\begin{equation}
R_{M}^{({\rm diff.})}\left(s,\,y,\,p_{T}\right)\equiv\frac{d\sigma_{M}^{{\rm diff}}/dy\,dp_{T}}{d\sigma_{M}^{{\rm incl}}/dy\,dp_{T}},\quad M=D^{\pm},B^{\pm},\,...,\label{eq:RDiff-2}
\end{equation}
which presents a novel observable. In Figure~\ref{fig:pTDependenceRatio}
we show this ratio as a function of $p_{T}$ for $D$-mesons, both
for prompt and non-prompt mechanisms. For the sake of definiteness
we considered $D^{+}$ mesons, although the results for the ratio~(\ref{eq:RDiff-2})
are almost the same for other choices of $D$-mesons. In Figure~\ref{fig:pTDependenceRatio-2}
we show the same ratio for the $B$-mesons ($B^{+}$ for definiteness)
and non-prompt $J/\psi$. We can see that the ratio is smaller than
for $D$-mesons, and decreases quite fast at large $p_{T}$. This
behavior agrees with our earlier observation that the single-diffractive
mechanism is formally a higher twist effect compared to the dominant
two-gluon fusion mechanism, in the case of inclusive production. As
expected, at small $p_{T}$ the ratios are similar for $B$-mesons
and non-prompt $J/\psi$; for larger $p_{T}$ the results differ due
to differences in fragmentation functions (see Appendix~\ref{sec:FragFunctions}
for details).

\begin{figure}
\includegraphics[width=9cm]{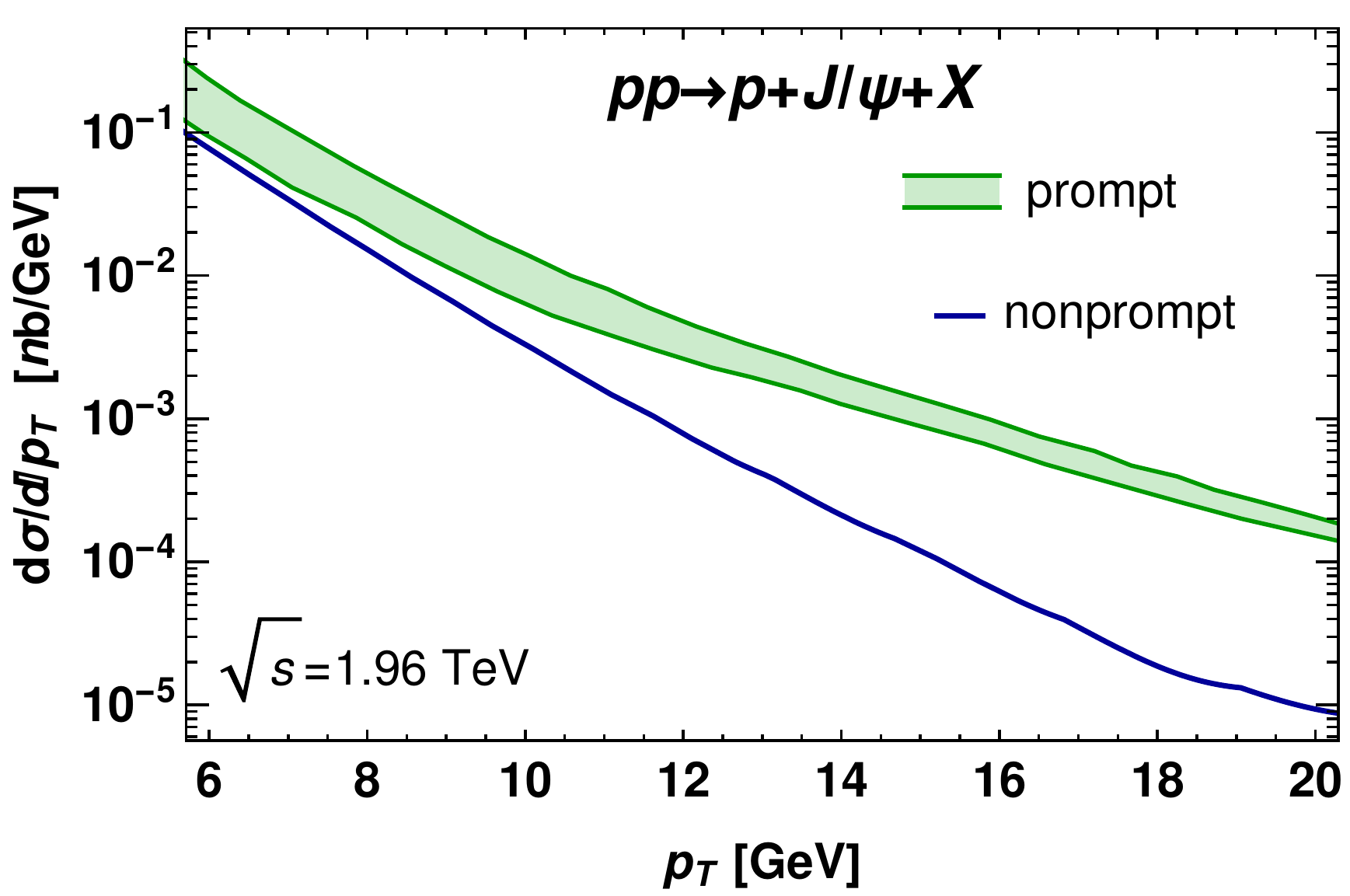}\includegraphics[width=9cm]{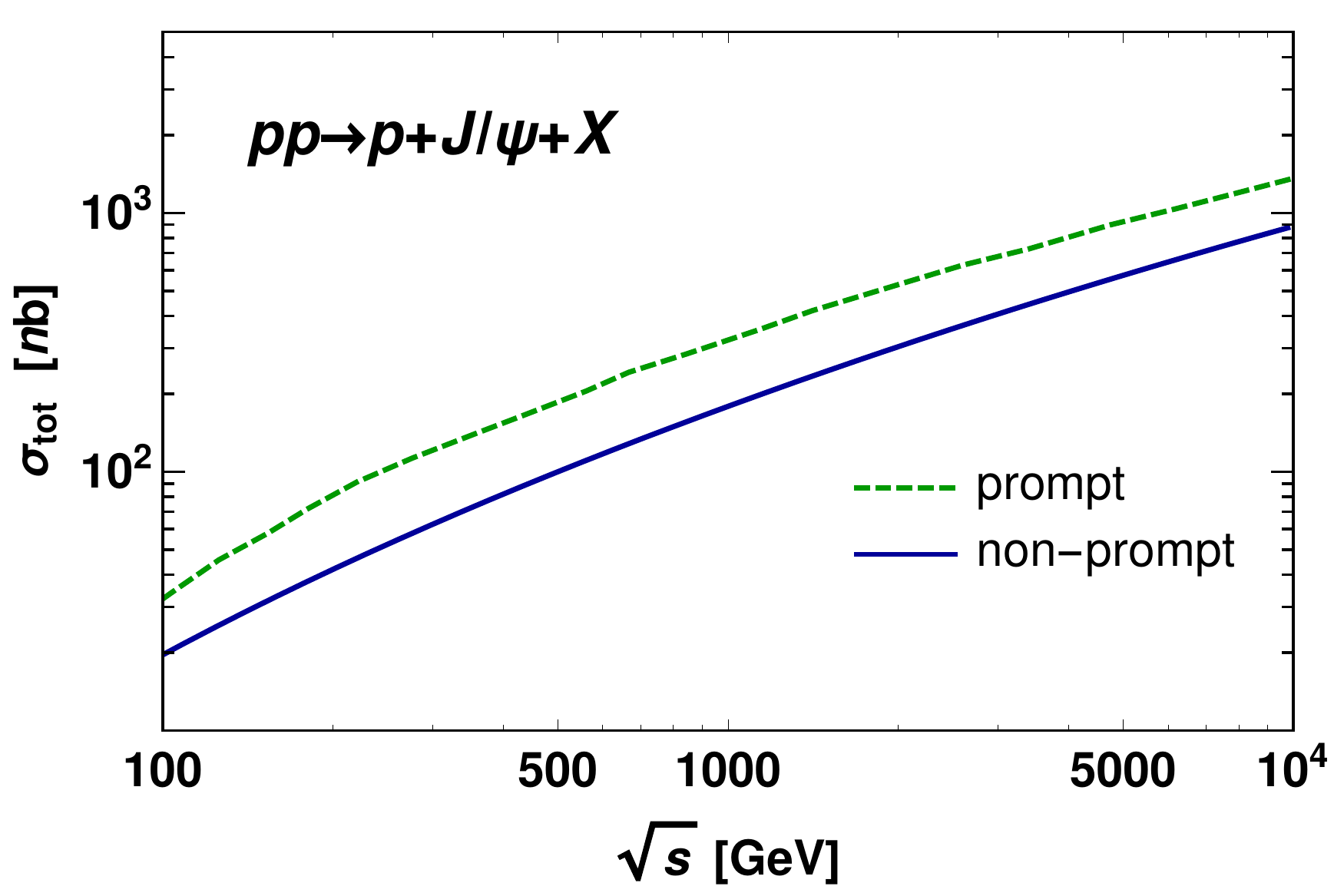}

\caption{\label{fig:pTDependence-promptVSnonprompt}Left plot: $p_{T}$-dependence
of differential cross-sections of prompt and non-prompt mechanisms
for single diffractive production of $J/\psi$ mesons. The results
for the prompt mechanism are taken from~\cite{Yuan:1998rq,Yuan:1998qw},
and the width of the green band reflects the uncertainty due to one
of the model parameters (gluon fraction of pomeron $f_{g}$). The
results for the non-prompt mechanism (blue solid curve) are results
of this paper. Right plot: Energy dependence of total cross-sections
of prompt and non-prompt single diffractive production mechanisms
of $J/\psi$-mesons. The prompt contribution (green dashed line) is
taken from~\cite{Machado:2007vw}. }
\label{Diags_JMesons-2-1} 
\end{figure}

In Figure~\ref{fig:pTDependence-promptVSnonprompt} we compare our
results for non-prompt production of $J/\psi$ with the predictions
for \emph{prompt} production from~\cite{Yuan:1998rq,Yuan:1998qw}
(color octet contributions + gluon fragmentation, dominant at large
$p_{T}$) and from~\cite{Machado:2007vw} (color evaporation model).
As we can expect, the non-prompt mechanism is smaller than the prompt
contribution, although the qualitative behavior is similar in both
cases.

\begin{figure}
\includegraphics[width=9cm]{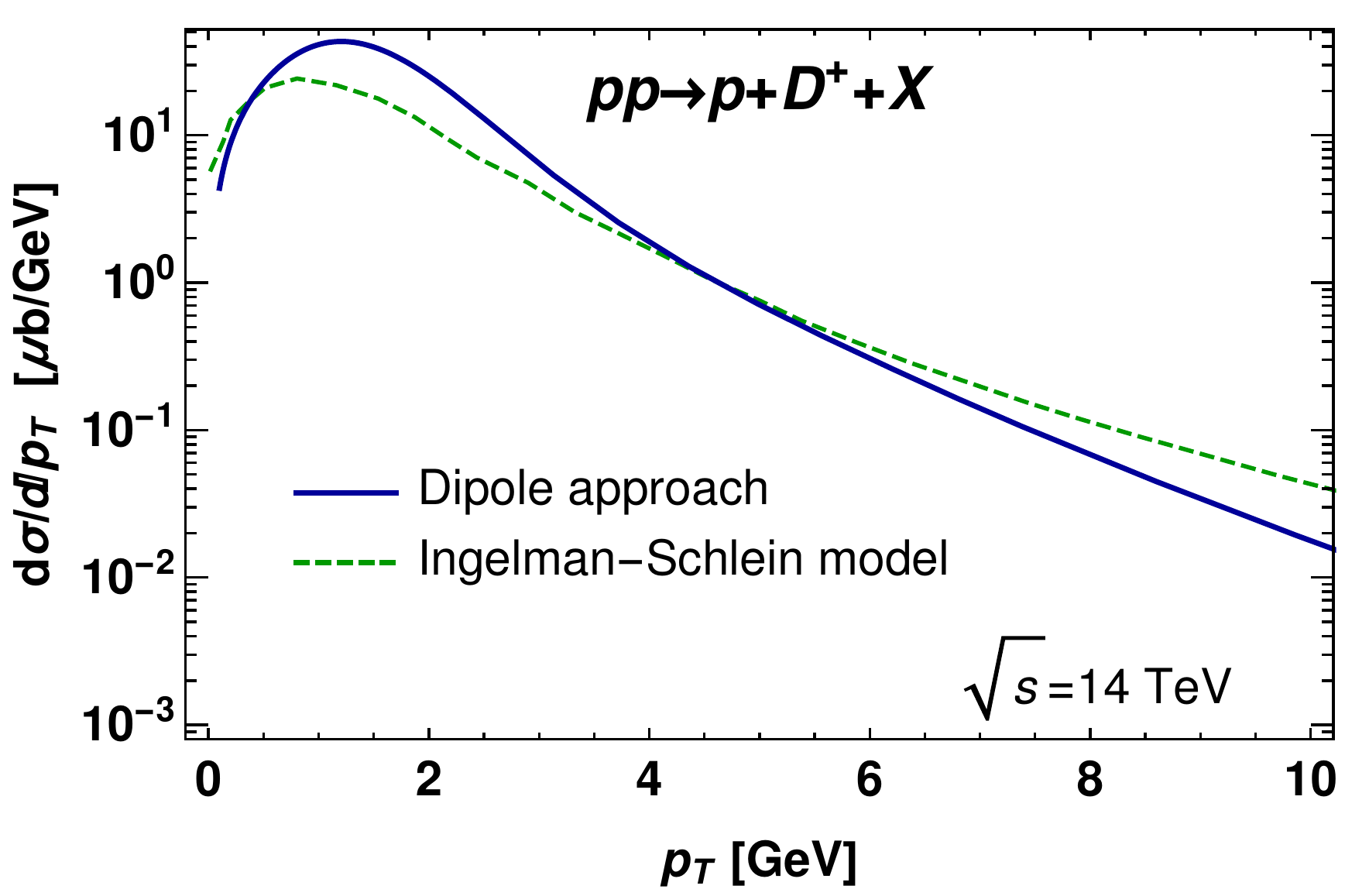}\includegraphics[width=9cm]{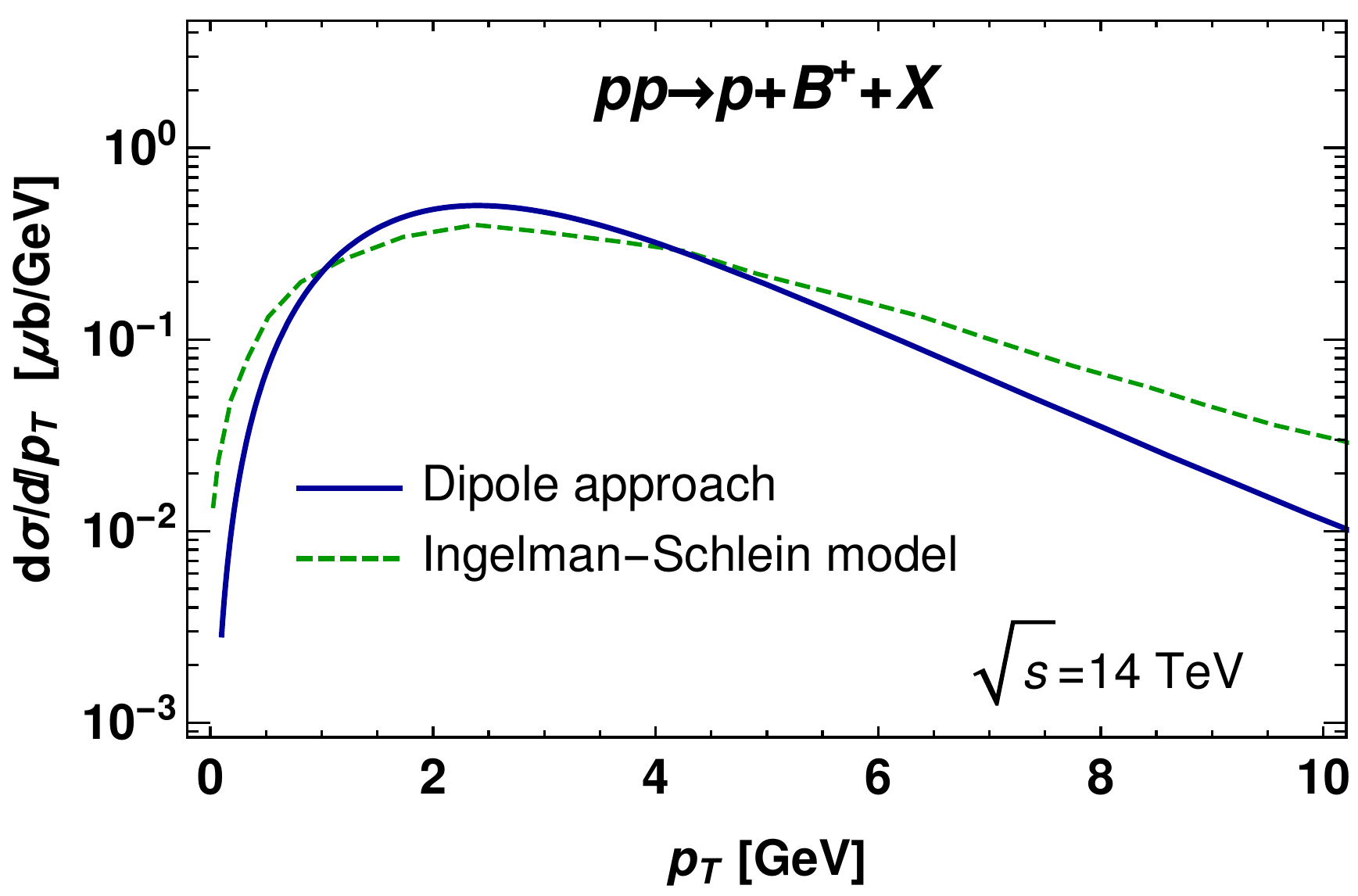}

\caption{\label{fig:pTDependence-ComparisonWithSzczurek}Comparison of color
dipole approach predictions (this paper) with results of~\cite{Luszczak:2014cxa}
obtained in the framework of Ingleman-Schlein model~\cite{Ingelman:1984ns}.
The left plot corresponds to single-diffractive charm production,
the right plot is for bottom quarks. }
\label{Diags_JMesons-2-1-2} 
\end{figure}

In Figure~\ref{fig:pTDependence-ComparisonWithSzczurek} we compare
our predictions with earlier results from~\cite{Luszczak:2014cxa}
obtained in the framework of Ingelman-Schlein model. We can see that
in the region $p_{T}\lesssim$5 GeV, where a majority of heavy mesons
are produced, both approaches give comparable contributions. At larger
$p_{T}$ the discrepancy between the two approaches increases.

Finally, we would like to stop briefly on the ratio $R_{J/\psi}^{({\rm diff})}$
of single diffractive and inclusive contributions. It was predicted
in~\cite{Yuan:1998qw} that for the prompt contributions $R_{J/\psi}^{({\rm diff,\,prompt})}\approx0.65\pm0.15\%$,
although later the CDF collaboration~\cite{Affolder:2001nc} found
a value twice larger 
\begin{equation}
R_{J/\psi}^{({\rm diff,\,CDF})}\approx1.45\pm0.25\%\label{eq:R_CDF}
\end{equation}
This mismatch might be explained by sizeable non-prompt contributions:
combining $R_{J/\psi}^{({\rm diff,\,prompt})}$ with $R_{J/\psi}^{({\rm diff,\,non-prompt})}$
from the first line in Table \ref{tab:Ratios}, we get $R_{J/\psi}^{({\rm diff,\,prompt+nonprompt})}\approx1.22\%$,
in reasonable agreement with the experimental value~(\ref{eq:R_CDF}).

\section{Multiplicity dependence}

\label{sec:MultiplicityGeneralities}According to the Local Parton
Hadron Duality (LPHD) hypothesis~\cite{LPHD1,LPHD2,LPHD3}, the multiplicity
of produced hadrons in a given event is directly related to the number
of partons produced in a collision. For this reason the study of multiplicity
dependence of different processes presents an interesting extension,
which allows to understand better the onset of the saturation regime
in high energy collisions. A feasibility to measure such processes
was demonstrated for inclusive channels by the STAR~\cite{Ma:2015xta,Trzeciak:2015fgz}
and ALICE~\cite{Adam:2015ota,Abelev:2012rz} collaborations. The
extension of these experimental measurements to single diffractive
production is quite straightforward, since their detectors have the
capability to detect simultaneously both the rapidity gaps and the
charged particles outside of the rapidity window. Since the cross-section
of single diffractive production is significantly smaller than that
of inclusive production, and the probability of events with large
multiplicity is exponentially suppressed~\cite{Abelev:2012rz}, each
measurement will require larger integrated luminosity.

In order to get rid of a common exponential suppression at large multiplicities,
for a comparison of the multiplicity dependence in different channels
it is widely accepted accepted to use a self-normalized ratio~\cite{Thakur:2018dmp}
\begin{align}
\frac{dN_{M}/dy}{\langle dN_{M}/dy\rangle}\,\,=\frac{w\left(N_{M}\right)}{\left\langle w\left(N_{M}\right)\right\rangle }\,\frac{\left\langle w\left(N_{{\rm ch}}\right)\right\rangle }{w\left(N_{{\rm ch}}\right)}= & \frac{d\sigma_{M}\left(y,\,\eta,\,\sqrt{s},\,n\right)/dy}{d\sigma_{M}\left(y,\,\eta,\,\sqrt{s},\,\langle n\rangle=1\right)/dy}\bigg/\frac{d\sigma_{{\rm ch}}\left(\eta,\,\sqrt{s},\,Q^{2},\,n\right)/d\eta}{d\sigma_{{\rm ch}}\left(\eta,\,\sqrt{s},\,Q^{2},\,\langle n\rangle=1\right)/d\eta}\label{eq:NDef}
\end{align}
where $\langle N_{{\rm ch}}\rangle=\Delta\eta\,dN_{{\rm ch}}/d\eta$
is the average number of particles detected in a given pseudorapidity
window $(\eta-\Delta\eta/2,\,\,\eta+\Delta\eta/2)$, $n=N_{{\rm ch}}/\langle N_{{\rm ch}}\rangle$
is the relative enhancement of the number of charged particles in
the same pseudorapidity window, $w\left(N_{M}\right)/\left\langle w\left(N_{M}\right)\right\rangle $
and $w\left(N_{{\rm ch}}\right)/\left\langle w\left(N_{{\rm ch}}\right)\right\rangle $
are the self-normalized yields of heavy meson $M$ ($M=D,\,B$) and
charged particles (minimal bias events) in a given multiplicity class;
$d\sigma_{M}(y,\,\sqrt{s},\,n)$ is the production cross-sections
for heavy meson $M$ with rapidity $y$ and $N_{{\rm ch}}=n\,\langle N_{{\rm ch}}\rangle$
charged particles in the pseudorapidity window $(\eta-\Delta\eta/2,\,\,\eta+\Delta\eta/2)$,
whereas $d\sigma_{{\rm ch}}(y,\,\sqrt{s},\,n)$ is the production
cross-sections for $N_{{\rm ch}}=n\,\langle N_{{\rm ch}}\rangle$
charged particles in the same pseudorapidity window. Mathematically
the ratio~(\ref{eq:NDef}) gives a \emph{conditional} probability
to produce a meson $M$ in a single diffractive collision in which
$N_{{\rm ch}}$ charged particles are produced.

In the color dipole (CGC/Sat) approach, the framework for description
of the high-multiplicity events has been developed in~\cite{KOLEB,KLN,DKLN,Kharzeev:2000ph,Kovchegov:2000hz,LERE,Lappi:2011gu,Ma:2018bax}.
In this picture the observation of enhanced multiplicity signals that
a larger than average number of partons is produced in a given event.
Nevertheless, we still expect that each pomeron should satisfy the
nonlinear Balitsky-Kovchegov equation. The bCGC dipole amplitude~(\ref{eq:CGCDipoleParametrization})
was constructed as an approximate solution of the latter, and for
this reason it should maintain its form, although the value of the
saturation scale $Q_{s}$ might be modified. As was demonstrated in~\cite{KOLEB,KLN,DKLN},
the observed number of charged multiplicity $dN_{{\rm ch}}/dy$ of
soft hadrons in $pp$ collisions is proportional to the saturation
scale $Q_{s}^{2}$ (modulo logarithmic corrections), for this reason
the events with large multiplicity might be described in dipole framework
by simply rescaling $Q_{s}^{2}$ as a function of $n$~\cite{KOLEB,KLN,DKLN,Kharzeev:2000ph,Kovchegov:2000hz,LERE,Lappi:2011gu},

\begin{equation}
Q_{s}^{2}\left(x,\,b;\,n\right)\,\,=\,\,n\,Q^{2}\left(x,\,b\right).\label{QSN-1}
\end{equation}
It was demonstrated in~\cite{Ma:2018bax} that the error of the approximation~~(\ref{QSN-1})
is less than 10\% in the region of interest ($n\lesssim10$), and
for this reason we will use it for our estimates. While at LHC energies
it is expected that the typical values of saturation scale $Q_{s}\left(x,\,b\right)$
fall into the range 0.5-1 ${\rm GeV}$, from~(\ref{QSN-1}) we can
see that in events with enhanced multiplicity this parameter might
exceed the values of heavy quark mass $m_{Q}$ and lead to an interplay
of large-$Q_{s}$ and large-$m_{Q}$ limits. The expression~(\ref{QSN-1})
explicitly illustrates that the study of the high-multiplicity events
gives us access to a new regime, which otherwise would require significantly
higher energies.

\begin{figure}
\includegraphics[width=9cm]{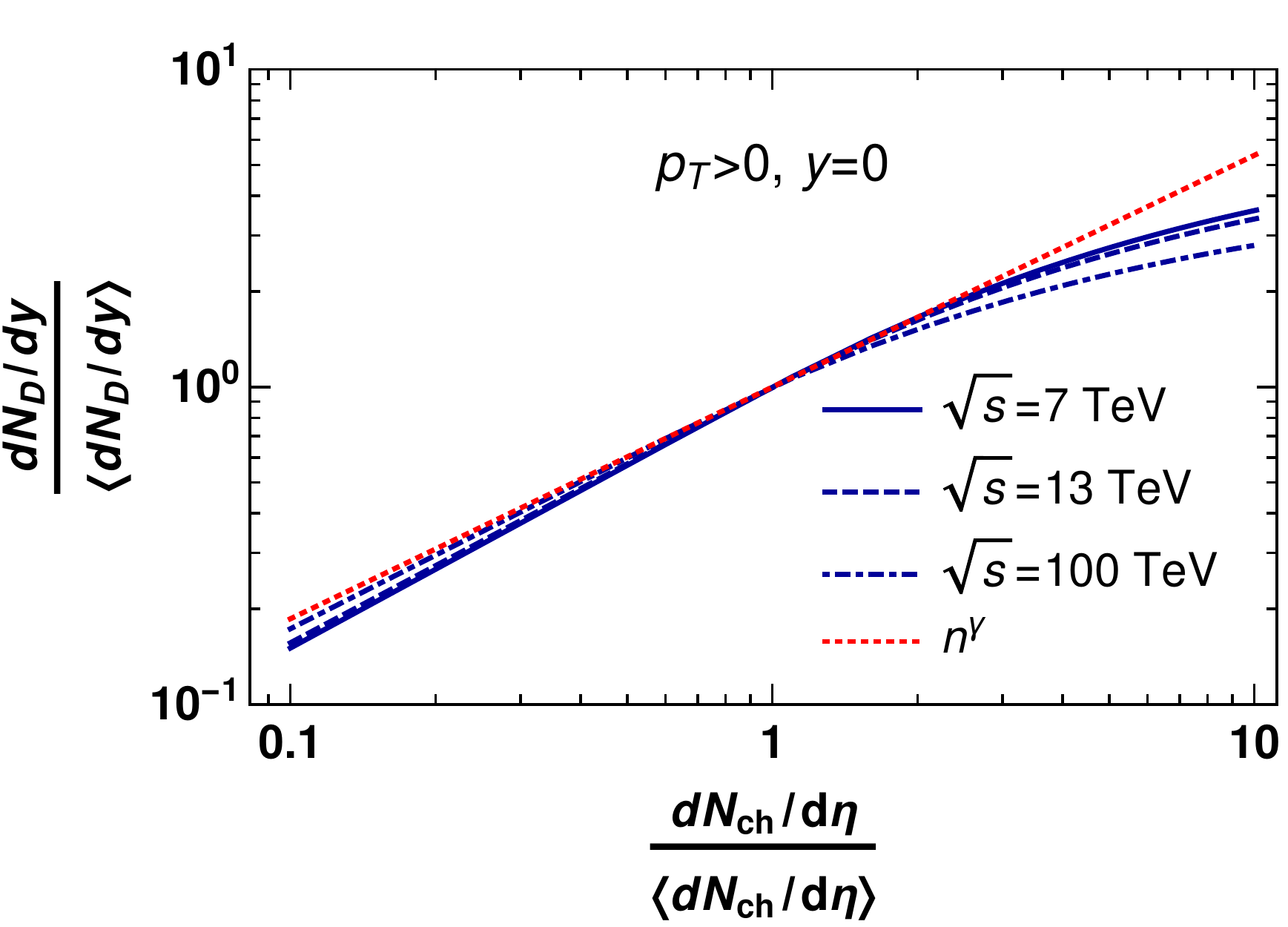}\includegraphics[width=9cm]{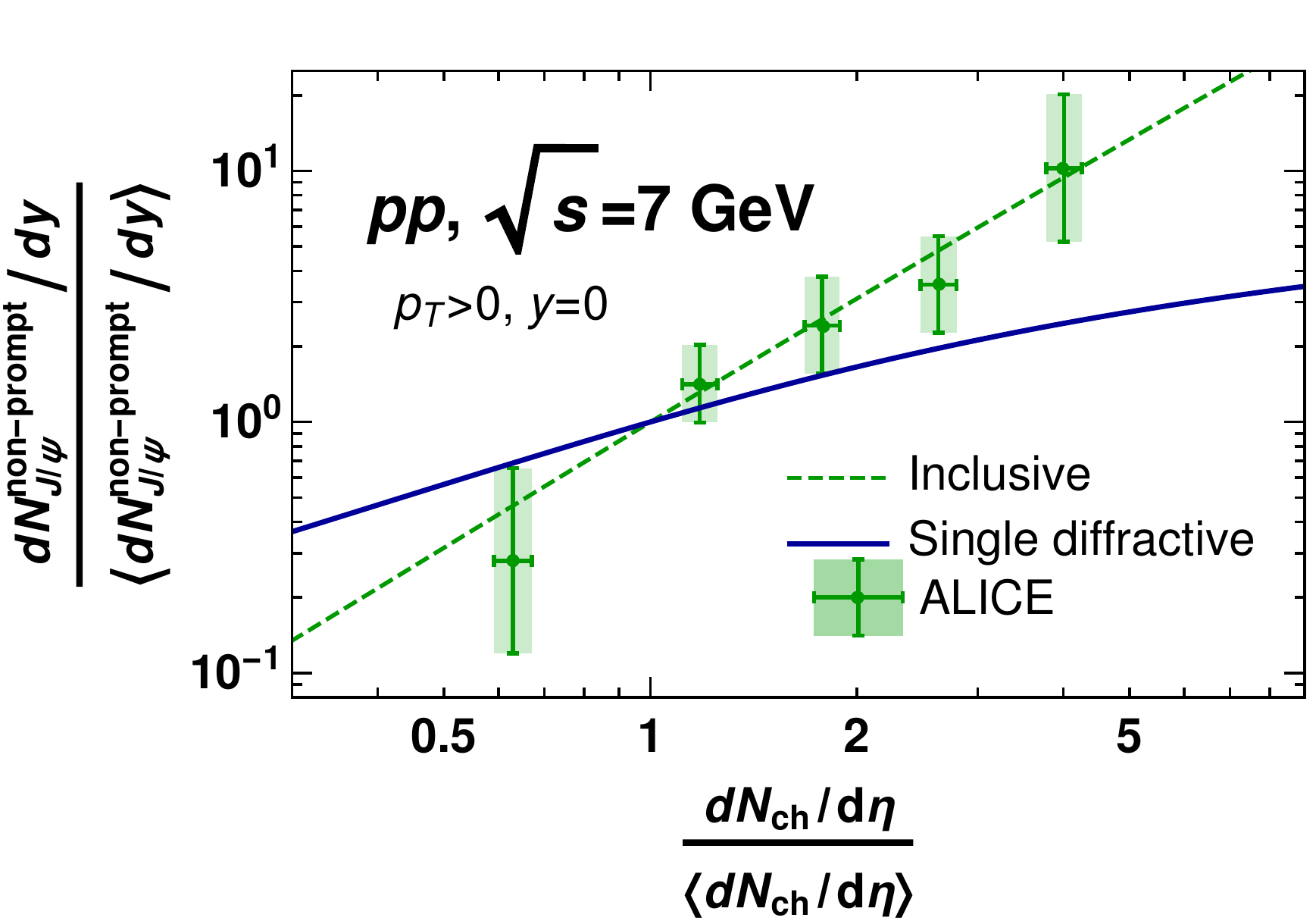}

\caption{\label{fig:MultiplicityDependence}Left plot: Multiplicity dependence
of open heavy flavor meson production cross-sections with single diffractive
mechanism (the same for all mesons, see the text for explanation).
The red dotted line corresponds to the asymptotic expression for small
multiplicities, as explained in the text. Right plot: comparison of
multiplicity dependence for inclusive and single diffractive production
for non-prompt $J/\psi$ mesons. The experimental points are from
ALICE~\cite{Adam:2015ota} for inclusive production, the theoretical
curve for inclusive production is from~\cite{Schmidt:2020fgn}.}
\label{Diags_JMesons-2-1-1} 
\end{figure}

The observation of enhanced multiplicity in the process shown in the
left diagram of Figure~\ref{fig:SD} implies that unintegrated gluon
density $g\left(x,\,\boldsymbol{k}_{\perp},\,n\right)$ in~(\ref{FD1-2})
is also modified. This change might be found taking into account the
relation of gluon density with the dipole amplitude $N(x,\,r,\,b)$
given by~(\ref{eq:GN3-1}). For the sake of simplicity below we'll
focus on the multiplicity dependence of the $p_{T}$-integrated cross-section,
which is easier to measure experimentally. For this case the cross-section~(\ref{FD1-2})
simplifies considerably, since, after integration over $p_{T}$, the
multiplicity dependent (integrated) gluon density factorizes and contributes
to the result as a multiplicative factor. For this reason the ratio~(\ref{eq:NDef})
reduces to a common factor 
\begin{equation}
\frac{dN_{M}/dy}{\langle dN_{M}/dy\rangle}\,\,=\frac{\int d^{2}r\,\frac{J_{1}\left(r\,\mu_{F}\right)}{r}\nabla_{r}^{2}N\left(y,\,\boldsymbol{r},\,n\right)}{\int d^{2}r\,\frac{J_{1}\left(r\,\mu_{F}\right)}{r}\nabla_{r}^{2}N\left(y,\,\boldsymbol{r},\,1\right)},\label{eq:NDef-1}
\end{equation}
the same for all mesons. In Figure~\ref{fig:MultiplicityDependence}
we show the multiplicity dependence of the ratio~(\ref{eq:NDef-1}).
At very small $n$, when saturation effects are small, the size of
the dipole is controlled by the mass of heavy quark $\sim1/m_{Q}$,
and thus the dipole amplitude $N\left(y,\,\boldsymbol{r},\,n\right)$
might be approximated as $N\left(y,\,\boldsymbol{r},\,n\right)\sim\left(r\,Q_{s}\left(y,\,n\right)\right)^{\gamma}$,
where $\gamma\approx0.63-0.76$ is a numerical parameter. In view
of (\ref{QSN-1}) this translates into the multiplicity dependence
\begin{equation}
\frac{dN_{M}/dy}{\langle dN_{M}/dy\rangle}\sim n^{\gamma},
\end{equation}
as shown in the same Figure~\ref{fig:MultiplicityDependence} with
red dotted line. At larger values of $n$, due to saturation effects,
the curve deviates from the small-$n$ asymptotic behavior. As we
can see from the right panel of the same Figure~\ref{fig:MultiplicityDependence},
this behavior is different from the dependence seen by ALICE for \emph{inclusive}
the production~\cite{Adam:2015ota}, as well as from our theoretical
result for inclusive production from~\cite{Schmidt:2020fgn}. This
happens because in single diffractive production the co-produced hadrons
stem from only one cut pomeron, whereas in inclusive production, in
the setup studied in~\cite{Adam:2015ota}, at least two pomerons
can contribute to the observed multiplicity enhancement. Since each
cut pomeron gives a factor $\sim n^{\gamma}$ in multiplicity dependence,
this explains the predicted difference between the single diffractive
and inclusive processes.

\section{Nuclear effects}

\label{sec:Nucl}The study of the single diffractive production on
nuclear collisions is appealing because its cross-section grows rapidly
with atomic number $A$, and thus is easier to measure experimentally.
The $AA$ collisions are not suitable for this purpose due to formation
of hot Quark-Gluon Plasma at later stages~\cite{Shuryak:1978ij,Shuryak:1980tp,BraunMunzinger:2007zz,Fukushima:2011nq,Bjorken:1982qr,Schmidt:2018rkw,Collins:1974ky}.
For this reason we will focus on $pA$ collisions and in the kinematics
when the scattered proton in the final state is separated by large
rapidity gap from the produced heavy meson and nuclear debris. 

\begin{figure}
\includegraphics[width=9cm]{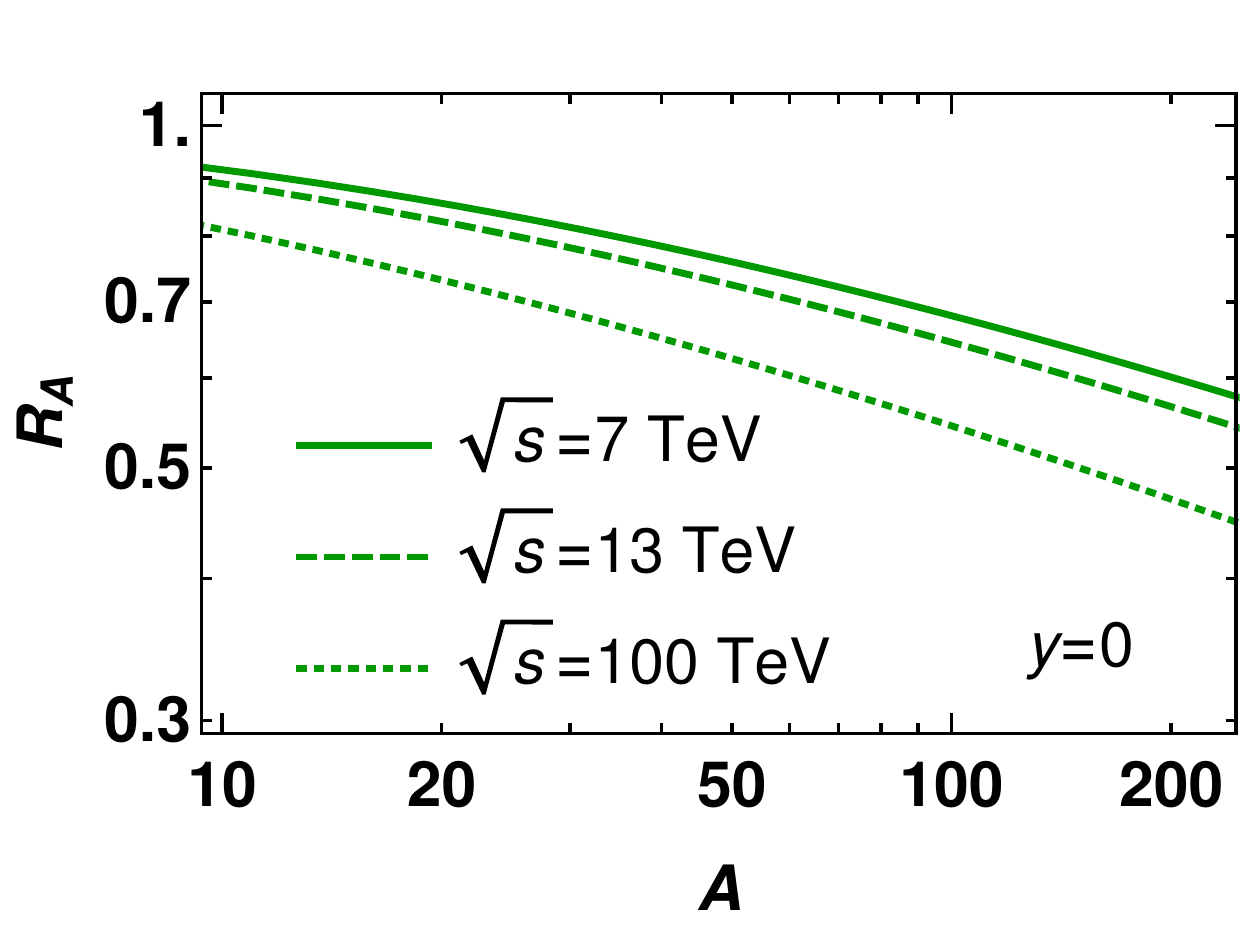}

\caption{\label{fig:ADependence}The nuclear suppression factor~$R_{A}$ defined
in~(\ref{eq:RatioNucl}) as a function of the atomic number $A$
for the $p_{T}$-integrated cross-section (the same for all mesons,
see the text for explanation). }
\label{Diags_JMesons-2-1-1-1} 
\end{figure}

In CGC framework the nucleus differs from the proton by larger size
$R_{A}=A^{1/3}R_{p}$ and larger values of the saturation scale $Q_{sA}^{2}$.
As was found in~\cite{Albacete:2004gw,Albacete:2005ef} from analysis
of the experimental data, the dependence of $Q_{sA}^{2}$ on atomic
number $A$ might be approximated by 
\begin{equation}
Q_{sA}^{2}(x)\approx Q_{s}^{2}(x)\,A^{1/3\delta}\qquad\delta\approx0.79\pm0.02.\label{eq:SatNUcl}
\end{equation}
The value $\delta<1$ indicates that the saturation scale grows \emph{faster}
than $\sim A^{1/3}$ expected from naive geometric estimates. In single
diffractive process  the nucleus contributes in~(\ref{FD1-2}) only
through the unintegrated gluon density $g(x,\,\boldsymbol{k})$. Currently
the latter is poorly defined experimentally~~\cite{Eskola:2009uj},
for this reason we will estimate it from the dipole amplitude using~(\ref{GN2-1},\ref{eq:GN3-1}).
The magnitude of nuclear effects is conventionally expressed in terms
of the normalized ratio of the cross-sections on the nucleus and proton,
\begin{equation}
R_{A}(y)=\frac{d\sigma_{pA\to p\,M\,X}/dy}{A\,d\sigma_{pp\to p\,M\,X}/dy}.\label{eq:RatioNucl}
\end{equation}
For the sake of simplicity we'll focus on the $p_{T}$-integrated
cross-section. In this case the dependence on the gluon PDF factorizes,
and thus the ratio~(\ref{eq:RatioNucl}) reduces to a common prefactor 

\begin{align}
R_{A}(y) & \approx\frac{g_{A}\left(x_{1}(y),\,\mu_{F}\right)}{g_{N}\left(x_{1}(y),\,\mu_{F}\right)}=\frac{1}{A}\frac{\int d^{2}b\int d^{2}r\,\frac{J_{1}\left(r\,\mu_{F}\right)}{r}\nabla_{r}^{2}N_{A}\left(y,\,\boldsymbol{r},\,\boldsymbol{b}/A^{1/3}\right)}{\int d^{2}b\int d^{2}r\,\frac{J_{1}\left(r\,\mu_{F}\right)}{r}\nabla_{r}^{2}N\left(y,\,\boldsymbol{r},\,\boldsymbol{b}\right)},\label{eq:NDef-A}
\end{align}
where $N_{A}\left(y,\,\boldsymbol{r},\,\boldsymbol{b}\right)$ is
a nuclear dipole amplitude with adjusted saturation scale~(\ref{eq:SatNUcl}),
and the rescaling of the impact parameter $\boldsymbol{b}$ in the
numerator reflects the increase of the nuclear radius. In the Figure~\ref{fig:ADependence}
we have shown the ratio~(\ref{eq:RatioNucl}) as a function of the
atomic number $A$. We can see that due to nuclear (saturation) effects
the cross-section decreases by up to a factor of two for very heavy
nuclei. This finding is in agreement with expected suppression of
nuclear gluon densities found in~\cite{Eskola:2009uj} from global
fits of experimental data.

Finally, from comparison of~(\ref{eq:NDef-1}) and (\ref{eq:NDef-A})
we may obtain the relation between the \emph{nuclear} suppression
factor $R_{A}$ and the multiplicity dependence of the \emph{proton}
cross-section~(\ref{eq:NDef-1}),
\[
A\,R_{A}(y,\,A)=\left.\frac{dN_{M}/dy}{\langle dN_{M}/dy\rangle}\right|_{n=\left(Q_{sA}^{2}/Q_{s}^{2}\right)},
\]
which might be checked experimentally.

\section{Conclusions}

\label{sec:Conclusions}In this paper we studied single diffractive
production of open heavy-flavor mesons. We analyzed in detail the
production of $D$- and $B$-mesons, as well as non-prompt production
of $J/\psi$ mesons. While in general diffractive events constitute
up to 20 per cent of inclusive cross-section~\cite{Abelev:2012sea},
we found that for heavy mesons production the single diffractive events
constitutes only 0.4-2 per cent of all inclusively produced heavy
mesons. This happens because the leading order contribution to single
diffractive production is formally a higher twist effect (compared
to leading order inclusive diagrams) and thus includes additional
suppression $\sim\left(\Lambda_{{\rm QCD}}/m_{Q}\right)^{2}$. Similarly,
the observed suppression at large transverse momentum $p_{T}$ of
the produced heavy meson agrees with expected pattern of higher twist
suppression. Nevertheless, we believe that the cross-sections are
sufficiently large and thus could be measured with reasonable precision
at the LHC.

We also analyzed the dependence on multiplicity of co-produced hadrons,
assuming that these are produced only on one side of the heavy meson.
We found that the dependence on multiplicity is mild, in contrast
to the vigorously growing multiplicity seen by ALICE~\cite{Adam:2015ota}
for inclusive production. Our evaluation is largely parameter-free
and relies only on the choice of the parametrization for the dipole
cross-section~(\ref{eq:CGCDipoleParametrization}).

We expect that suggested processes might be studied by the CMS~ (see
their recent feasibility study in \cite{CMS:2014rga}), ALICE~\cite{Adam:2015ota,Abelev:2012rz}
and STAR collaborations.

\section*{Acknowldgements}

We thank our colleagues at UTFSM university for encouraging discussions.
This research was partially supported by the project Proyecto Basal
FB 0821\emph{ }(Chile) and Fondecyt (Chile) grant 1180232. Also, we
thank Yuri Ivanov for technical support of the USM HPC cluster, where
some evaluations were performed.

\appendix

\section{Evaluation of the dipole amplitudes}

\label{sec:Derivation}

\subsection{Single diffractive production}

\label{subsec:SDDerivation} 
\begin{figure}
\includegraphics[width=9cm]{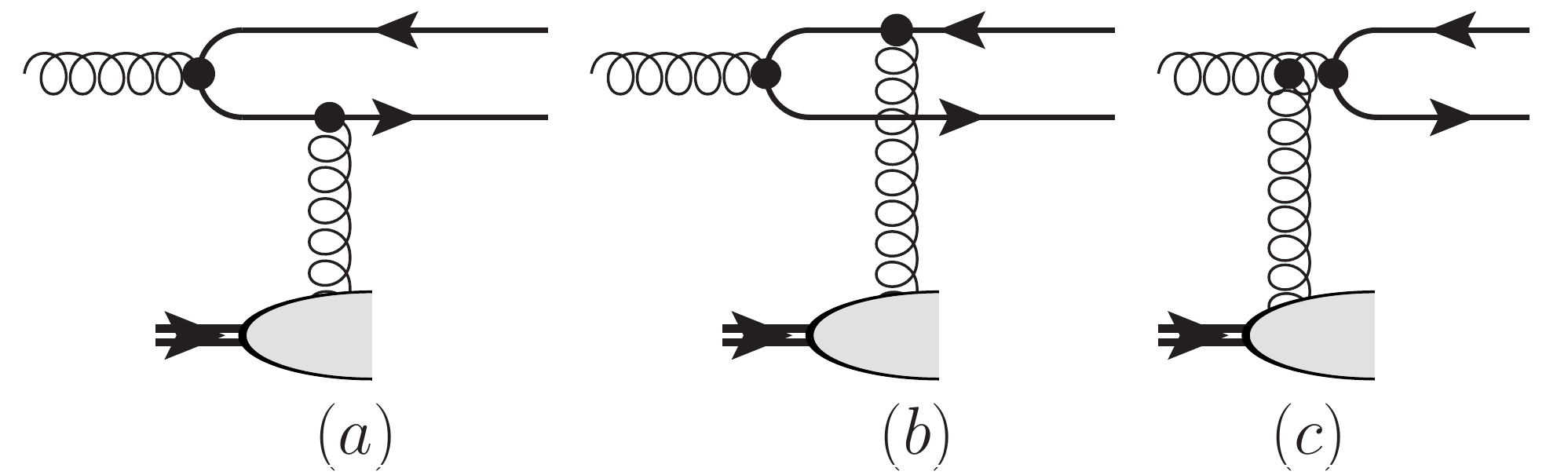}

\caption{\label{fig:Dipole2Pomeron}The diagrams which contribute to the heavy
meson production cross-section in the leading order perturbative QCD.
The contribution of the last diagram ($c$) to the meson formation
might be also viewed as gluon-gluon fusion $gg\to g$ with subsequent
gluon fragmentation $g\to\bar{Q}Q$. In CGC parametrization of the
dipole cross-section approach each ``gluon'' is replaced with reggeized
gluon (BK pomeron), which satisfies the Balitsky-Kovchegov equation
and corresponds to a fan-like shower of soft particles.}
\end{figure}

In this Appendix, for the sake of completeness, we explain the main
technical steps and assumptions used for the derivation of the single
diffractive cross-section~(\ref{FD1-2},~\ref{eq:N3Direct}). The
general rules which allow to express the cross-sections of hard processes
in terms of the color singlet dipole cross-section might be found
in~\cite{GLR,McLerran:1993ni,McLerran:1993ka,McLerran:1994vd,MUQI,MV,gbw01:1,Kopeliovich:2002yv,Kopeliovich:2001ee}.
In the heavy quark mass limit the strong coupling $\alpha_{s}(m_{Q})$
is small, which allows to consider the interaction of a heavy $\bar{Q}Q$
dipole with gluons perturbatively and discuss them similar to the
treatment of the $k_{T}$-factorization approach. At the same time
we tacitly assume that each such gluon should be understood as a parton
shower (``pomeron'').

In the high-energy eikonal picture, the interaction of the quarks
and antiquark with a $t$-channel gluon are described by a factor
$\pm ig\,t^{a}\gamma\left(\boldsymbol{x}_{\perp}\right)$, where $\boldsymbol{x}_{\perp}$
is the transverse coordinate of the quark, and the function $\gamma\left(\boldsymbol{x}_{\perp}\right)$
is related to a distribution of gluons in the target. This function
is related to a dipole cross-section $\sigma(x,\,\boldsymbol{r})$
as 
\begin{equation}
\Delta\sigma(x,\,\boldsymbol{r})\equiv\sigma(x,\,\infty)-\sigma(x,\,\boldsymbol{r})=\frac{1}{8}\int d^{2}b\left|\gamma\left(x,\,\boldsymbol{b}-z\boldsymbol{r}\right)-\gamma\left(x,\,\boldsymbol{b}+\bar{z}\boldsymbol{r}\right)\right|^{2}\label{eq:DipoleX}
\end{equation}
where $\boldsymbol{r}$ is the transverse size of the dipole, and
$z$ is the light-cone fraction of the dipole momentum carried by
the quarks. The equation~(\ref{eq:DipoleX}) might be rewritten in
the form 
\begin{equation}
\frac{1}{8}\int d^{2}\boldsymbol{b}\gamma(x,\,\boldsymbol{b})\gamma(x,\,\boldsymbol{b}+\boldsymbol{r})=\frac{1}{2}\sigma(x,\,\boldsymbol{r})+\underbrace{\int d^{2}b\,\left|\gamma(x,\,\boldsymbol{b})\right|^{2}-\frac{1}{2}\sigma(x,\,\infty)}_{={\rm const}}.\label{eq:SigmaDef}
\end{equation}
For very small dipoles, the dipole cross-section is related to the
gluon uPDF as~\footnote{In the literature definitions of the unintegrated PDF $\mathcal{F}\left(x,\,k_{\perp}\right)$
might differ by a factor $k_{\perp}^{2}$.} 
\begin{equation}
\sigma\left(x,\,\vec{\boldsymbol{r}}\right)=\frac{4\pi\alpha_{s}}{3}\int\frac{d^{2}k_{\perp}}{k_{\perp}^{2}}\mathcal{F}\left(x,\,k_{\perp}\right)\left(1-e^{ik\cdot r}\right)+\mathcal{O}\left(\frac{\Lambda_{{\rm QCD}}}{m_{c}}\right),\label{eq:Dip}
\end{equation}
so the functions $\gamma\left(x,\,\boldsymbol{r}\right)$ might be
also related to the unintegrated gluon densities. With the help of~(\ref{eq:SigmaDef}),
for many high energy processes it is possible to express the exclusive
amplitude or inclusive cross-section as a linear combination of the
\emph{color singlet} dipole cross-sections $\sigma(x,\,\boldsymbol{r})$
with different arguments. While in the deeply saturated regime we
can no longer speak about individual gluons (or pomerons), we expect
that the relations between the dipole amplitudes and color singlet
cross-sections should be valid even in this case.

\begin{figure}
\includegraphics[width=9cm]{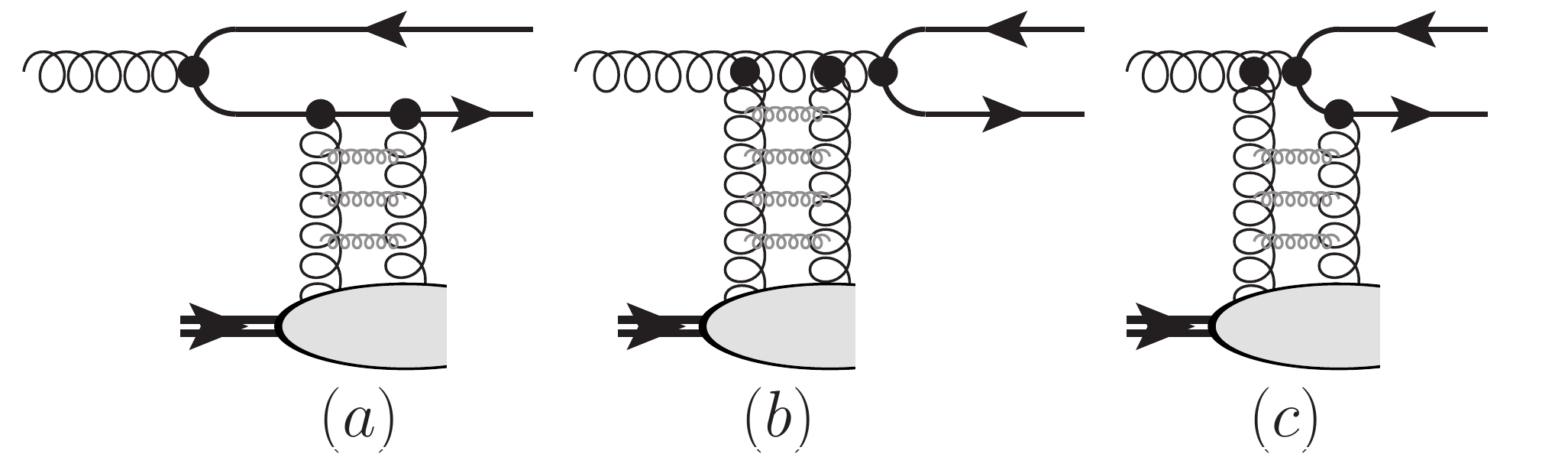}

\caption{\label{fig:Dipole3Pomeron}The diagrams which contribute to the single
diffractive heavy meson production in the leading order in perturbative
QCD ($\mathcal{O}(\alpha_{s})$-correction). In diagrams ($a$) and
$(c)$ all possible attachments of the gluon to the quarks and antiquarks
are implied. In QCD the interaction of the color dipole with a pomeron
might be understood as a gluon ladder (BFKL pomeron), for this reason
its interaction with a dipole is described as with a pair of gluons
in a color singlet state (see the text for explanation).}
\end{figure}

For the case of single-diffractive heavy quark pair production, the
leading-order contribution is given by the diagrams shown in the Figure~(\ref{fig:Dipole3Pomeron}).
As was explained at the beginning of this appendix, in the heavy quark
mass limit the interactions of $\bar{Q}Q$ with gluons become perturbative,
which implies that the $t$-channel pomeron might be considered as
a color singlet pair of gluons. Taking into account all the diagrams
shown in the Figure~\ref{fig:Dipole3Pomeron} and properties of the
$SU(N_{c})$ structure constants, we may express the amplitude of
the single diffractive process as

\begin{align*}
\mathcal{A}^{(3)}\left(x,\,\vec{\boldsymbol{r}}_{Q},\,\vec{\boldsymbol{r}}_{\bar{Q}}\right) & =\left[\frac{N_{c}}{4}\gamma_{+}^{2}\left(x,\,\vec{\boldsymbol{r}}_{Q},\,\vec{\boldsymbol{r}}_{\bar{Q}}\right)+\left(\frac{N_{c}^{2}-4}{4N_{c}}+\frac{1}{6}\right)\gamma_{-}^{2}\left(x,\,\vec{\boldsymbol{r}}_{Q},\,\vec{\boldsymbol{r}}_{\bar{Q}}\right)\right]t_{a}\\
 & \equiv a\left(x,\,\vec{\boldsymbol{r}}_{Q},\,\vec{\boldsymbol{r}}_{\bar{Q}}\right)t_{a}.
\end{align*}

where 
\begin{align*}
\gamma_{+}\left(x,\,\vec{\boldsymbol{r}}_{1},\,\vec{\boldsymbol{r}}_{2}\right) & =\gamma\left(x,\,\vec{\boldsymbol{r}}_{1}\right)+\gamma\left(x,\,\vec{\boldsymbol{r}}_{2}\right)-2\gamma\left(x,\,\frac{\vec{\boldsymbol{r}}_{1}+\vec{\boldsymbol{r}}_{2}}{2}\right),\\
\gamma_{-}\left(x,\,\vec{\boldsymbol{r}}_{1},\,\vec{\boldsymbol{r}}_{2}\right) & =\gamma\left(x,\,\vec{\boldsymbol{r}}_{1}\right)-\gamma\left(x,\,\vec{\boldsymbol{r}}_{2}\right),
\end{align*}
$a$ is the color index of the incident (projectile) gluon, and $\vec{\boldsymbol{r}}_{Q},\,\boldsymbol{r}_{\bar{Q}}$
are the coordinates of the quarks. For evaluation of the $p_{T}$-dependent
cross-section we need to project the coordinate space quark distribution
onto the state with definite transverse momentum $\boldsymbol{p}_{T}$,
so we have for the evaluate the additional convolution $\sim\int d^{2}r_{1}d^{2}r_{2}\,e^{ip_{T}\cdot\left(r_{1}-r_{2}\right)}$,
where $\vec{\boldsymbol{r}}_{1,2}$ are the coordinates of the quark
in the amplitude and its conjugate, viz: 
\begin{align}
\left|\mathcal{A}^{(3)}\left(\boldsymbol{p}_{T}\right)\right|^{2} & =\left(1+\eta^{2}\right)\int d^{2}\boldsymbol{x}_{\bar{Q}}\int d^{2}\boldsymbol{x}_{Q}\int d^{2}\boldsymbol{y}_{Q}\,e^{i\boldsymbol{p}_{T}\cdot\left(\boldsymbol{x}_{Q}-\boldsymbol{y}_{Q}\right)}\,\,\left.\left(\mathcal{A}^{(3)}\left(\vec{\boldsymbol{x}}_{i}\right)\right)^{*}\mathcal{A}^{(3)}\left(\vec{\boldsymbol{y}}_{i}\right)\right|_{\vec{\boldsymbol{x}}_{\bar{Q}}=\vec{\boldsymbol{y}}_{\bar{Q}}}\label{eq:A3Sq}\\
 & =\left(\frac{1+\eta^{2}}{2}\right)\int d^{2}\boldsymbol{x}_{\bar{Q}}\int d^{2}\boldsymbol{x}_{Q}\int d^{2}\boldsymbol{y}_{Q}\,e^{i\boldsymbol{p}_{T}\cdot\left(\boldsymbol{x}_{Q}-\boldsymbol{y}_{Q}\right)}a^{*}\left(x,\,\vec{\boldsymbol{x}}_{Q},\,\vec{\boldsymbol{x}}_{\bar{Q}}\right)a\left(x,\,\vec{\boldsymbol{y}}_{Q},\,\vec{\boldsymbol{x}}_{\bar{Q}}\right).\nonumber 
\end{align}
As discussed earlier, at high energies we may apply iteratively the
relation~(\ref{eq:DipoleX}) and express the three-pomeron dipole
amplitude in terms of the \emph{color singlet} dipole cross-sections,
as given in~(\ref{eq:N3Direct}). In the frame where the momentum
of the primordial gluon is not zero, we should take into account an
additional convolution with the momentum distribution of the incident
(``primordial'') gluons, as shown in~(\ref{FD1-2}), and was demonstrated
in~\cite{Goncalves:2017chx}.

\subsection{Inclusive production}

\label{subsec:InclusiveSummary}
\begin{figure}
\includegraphics[width=18cm]{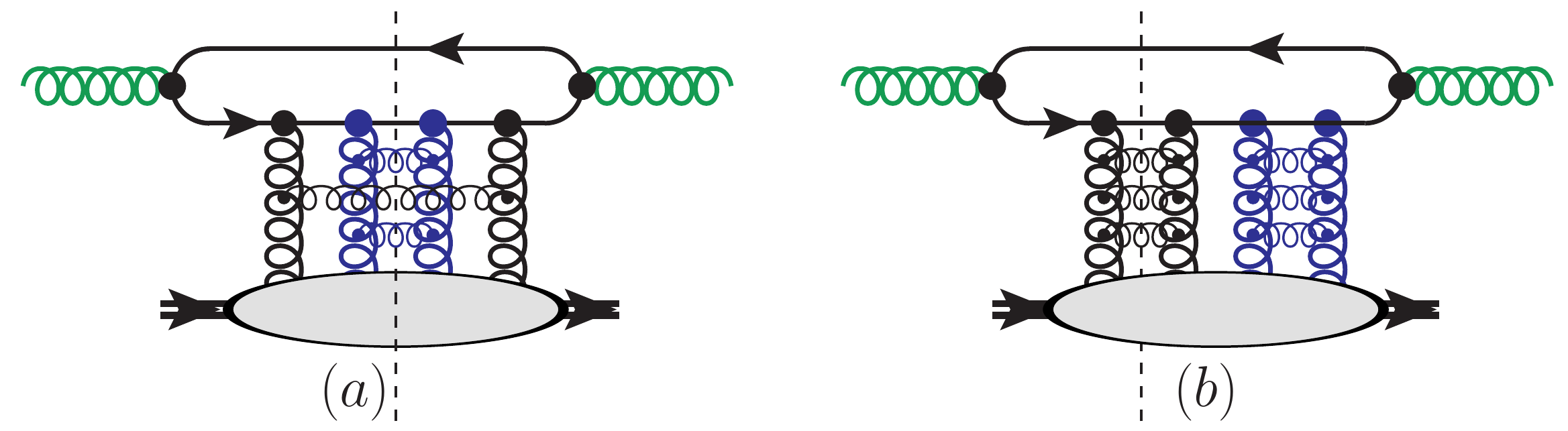}

\caption{\label{fig:NNLOInterference-CutUncutSummary} (color online) The three-pomeron
contributions (diagram ($a$)) contribute at the same order in $\alpha_{s}$
as the interference of LO and NNLO diagrams (diagram ($b$)). In both
plots the vertical dashed line is a unitary cut, lower blob is a target
(proton), and all possible connections of pomerons (thick wavy lines)
to the heavy $Q,\bar{Q}$ quark lines are implied. Note that in diagram
(a) both pomerons are cut, whereas in case of the interference contribution
one of the pomerons is uncut. }
\end{figure}

In Section~\ref{sec:Numer} we compared
predictions for single-diffractive production of heavy quarks with
those of the \emph{inclusive} production of the same mesons. For the
sake of completeness, in this Appendix we would like to mention briefly
the main expressions used for evaluation of the cross-sections for
the latter case. A detailed discussion of inclusive production, as
well as comparison with experimental data might be found in~\cite{Schmidt:2020fgn}.
The evaluation of the cross-section follows the steps outlined in
the previous Appendix~\ref{subsec:SDDerivation}. The leading order
contribution in the inclusive case is due to a standard fusion of
two gluons (pomerons). In the evaluation of the three-pomeron we should
take into account that there are two complementary mechanisms, shown
schematically in Figure~\ref{fig:NNLOInterference-CutUncutSummary}.
In what follows we'll refer to the contribution shown in the diagram
($a$) as genuine three-pomeron corrections, whereas the contribution
of the diagram ($b$) is the interference term. The two diagrams differ
by number of cut pomerons, and for this reason they have a different
multiplicity dependence. As we discussed in~\cite{Schmidt:2020fgn},
both twist-three corrections give sizeable contributions at small
$p_{T}\lesssim5$ GeV. For $D$-mesons the two corrections together
contribute up to 40-50 per cent of the leading order result, whereas
for $B$-mesons these contributions are of order 10\% even for $p_{T}\sim0$,
in agreement with the heavy mass limit.

Both the leading order cross-section and the higher twist correction
might be written as 
\begin{eqnarray}
 &  & \frac{d\sigma_{pp\to\bar{Q}_{i}Q_{i}+X}\left(y,\,\sqrt{s}\right)}{dy\,d^{2}p_{T}}=\,\int d^{2}k_{T}x_{1}\,g\left(x_{1},\,\boldsymbol{p}_{T}-\boldsymbol{k}_{T}\right)\int_{0}^{1}dz\int_{0}^{1}dz'\label{FD1-2-3}\\
 &  & \times\,\,\,\int\frac{d^{2}r_{1}}{4\pi}\,\int\frac{d^{2}r_{2}}{4\pi}e^{i\left(r_{1}-r_{2}\right)\cdot\boldsymbol{k}_{T}}\,\Psi_{\bar{Q}Q}^{\dagger}\left(r_{2},\,z,\,p_{T}\right)\Psi_{\bar{Q}Q}^{\dagger}\left(r_{1},\,z,\,p_{T}\right)\nonumber \\
 &  & \times N_{M}\left(x_{2}(y);\,\vec{r}_{1},\,\vec{r}_{2}\right)+\left(x_{1}\leftrightarrow x_{2}\right),\nonumber 
\end{eqnarray}
(see the Section~\ref{sec:Evaluation} for notations and definitions).
For the leading order contribution, the amplitude $N_{M}$ is given
by~\cite{Goncalves:2017chx,Schmidt:2020fgn} 
\begin{eqnarray}
 &  & N_{M}^{(2)}\left(x,\,\,\vec{r}_{1},\,\vec{r}_{2}\right)=\label{eq:N2}\\
 &  & =-\frac{1}{2}N\left(x,\,\vec{r}_{1}-\vec{r}_{2}\right)-\frac{1}{16}\left[N\left(x,\,\vec{r}_{1}\right)+N\left(x,\,\vec{r}_{2}\right)\right]-\frac{9}{8}N\left(x,\,\bar{z}\left(\vec{r}_{1}-\vec{r}_{2}\right)\right)\nonumber \\
 &  & +\frac{9}{16}\left[N\left(x,\,\bar{z}\vec{r}_{1}-\vec{r}_{2}\right)+N\left(x,\,\bar{z}\vec{r}_{2}-\vec{r}_{1}\right)+N\left(x,\,\bar{z}\vec{r}_{1}\right)+N\left(x,\,\bar{z}\vec{r}_{2}\right)\right].\nonumber 
\end{eqnarray}
Similarly, the three-pomeron contribution shown in the diagram $(a)$
of the Figure~\ref{fig:NNLOInterference-CutUncutSummary} may be
rewritten as

\begin{align}
N_{M}^{(3)}\left(x,\,z,\,\vec{\boldsymbol{r}}_{1},\,\vec{\boldsymbol{r}}_{2}\right)\approx & \,\frac{1}{8\sigma_{{\rm eff}}}\left[N_{+}^{2}\left(x,\,z,\,\vec{\boldsymbol{r}}_{1},\,\vec{\boldsymbol{r}}_{2}\right)\left(\frac{3N_{c}^{2}}{8}\right)+N_{-}^{2}\left(x,\,\vec{\boldsymbol{r}}_{1},\,\vec{\boldsymbol{r}}_{2}\right)\left(\frac{\left(43\,N_{c}^{4}-320N_{c}^{2}+720\right)}{72\,N_{c}^{2}}\right)\right.\label{eq:N3Direct-2}\\
 & \qquad+\left.\frac{\left(N_{c}^{2}-4\right)}{2}N_{+}\left(x,\,z,\,\vec{\boldsymbol{r}}_{1},\,\vec{\boldsymbol{r}}_{2}\right)N_{-}\left(x,\,\vec{\boldsymbol{r}}_{1},\,\vec{\boldsymbol{r}}_{2}\right)\right]\nonumber 
\end{align}
where 
\begin{align}
N_{-}\left(x,\,\vec{\boldsymbol{r}}_{1},\,\vec{\boldsymbol{r}}_{2}\right) & \equiv-\frac{1}{2}\left[N\left(x,\,\vec{\boldsymbol{r}}_{2}-\vec{\boldsymbol{r}}_{1}\right)-N\left(x,\,\vec{\boldsymbol{r}}_{1}\right)-N\left(x,\,\vec{\boldsymbol{r}}_{2}\right)\right]\\
N_{+}\left(x,\,z,\,\vec{\boldsymbol{r}}_{1},\,\vec{\boldsymbol{r}}_{2}\right) & \equiv-\frac{1}{2}\left[N\left(x,\,\vec{\boldsymbol{r}}_{2}-\vec{\boldsymbol{r}}_{1}\right)+N\left(x,\,\vec{\boldsymbol{r}}_{1}\right)+N\left(x,\,\vec{\boldsymbol{r}}_{2}\right)\right]+N\left(x,\,\bar{z}\vec{\boldsymbol{r}}_{1}-\vec{\boldsymbol{r}}_{2}\right)+N\left(x,\,\bar{z}\vec{\boldsymbol{r}}_{1}\right)\\
 & +N\left(x,\,-\bar{z}\vec{\boldsymbol{r}}_{2}+\vec{\boldsymbol{r}}_{1}\right)+N\left(x,\,-\bar{z}\vec{\boldsymbol{r}}_{2}\right)-2N\left(x,\,\bar{z}\left(\vec{\boldsymbol{r}}_{1}-\vec{\boldsymbol{r}}_{2}\right)\right)\nonumber 
\end{align}
and $\sigma_{{\rm eff}}\approx20\,{\rm mb}$ is a numerical parameter.
Finally, for the interference term shown in the diagram $(b)$ of
the Figure~ \ref{fig:NNLOInterference-CutUncutSummary} we may get
in a similar way

\begin{align}
N_{M}^{({\rm int})}\left(x,\,z,\,\vec{\boldsymbol{r}}_{1},\,\vec{\boldsymbol{r}}_{2}\right)= & -\,\frac{3}{16\,\sigma_{{\rm eff}}}\left[2\,N_{+}\left(x,\,z,\,\vec{\boldsymbol{r}}_{1},\,\vec{\boldsymbol{r}}_{2}\right)\tilde{N}_{+}\left(x,\,z,\,\vec{\boldsymbol{r}}_{2}\right)\left(\frac{3N_{c}^{2}}{8}\right)+\right.\label{eq:N3Interf}\\
 & -N_{-}\left(z,\,\vec{\boldsymbol{r}}_{1},\,\vec{\boldsymbol{r}}_{2}\right)\tilde{N}_{-}\left(x,\,\vec{\boldsymbol{r}}_{2}\right)\left(\frac{\left(43\,N_{c}^{4}-320N_{c}^{2}+720\right)}{72\,N_{c}^{2}}\right)+\nonumber \\
 & +\left.\frac{\left(N_{c}^{2}-4\right)}{2}\left(N_{+}\left(z,\,\vec{\boldsymbol{r}}_{1},\,\vec{\boldsymbol{r}}_{2}\right)\tilde{N}_{-}\left(x,\,\vec{\boldsymbol{r}}_{2}\right)+\tilde{N}_{+}\left(x,\,\vec{\boldsymbol{r}}_{2}\right)N_{-}\left(z,\,\vec{\boldsymbol{r}}_{1},\,\vec{\boldsymbol{r}}_{2}\right)\right)\right].\nonumber 
\end{align}

\section{Fragmentation functions}

\label{sec:FragFunctions} 
\begin{figure}
\includegraphics[width=9cm]{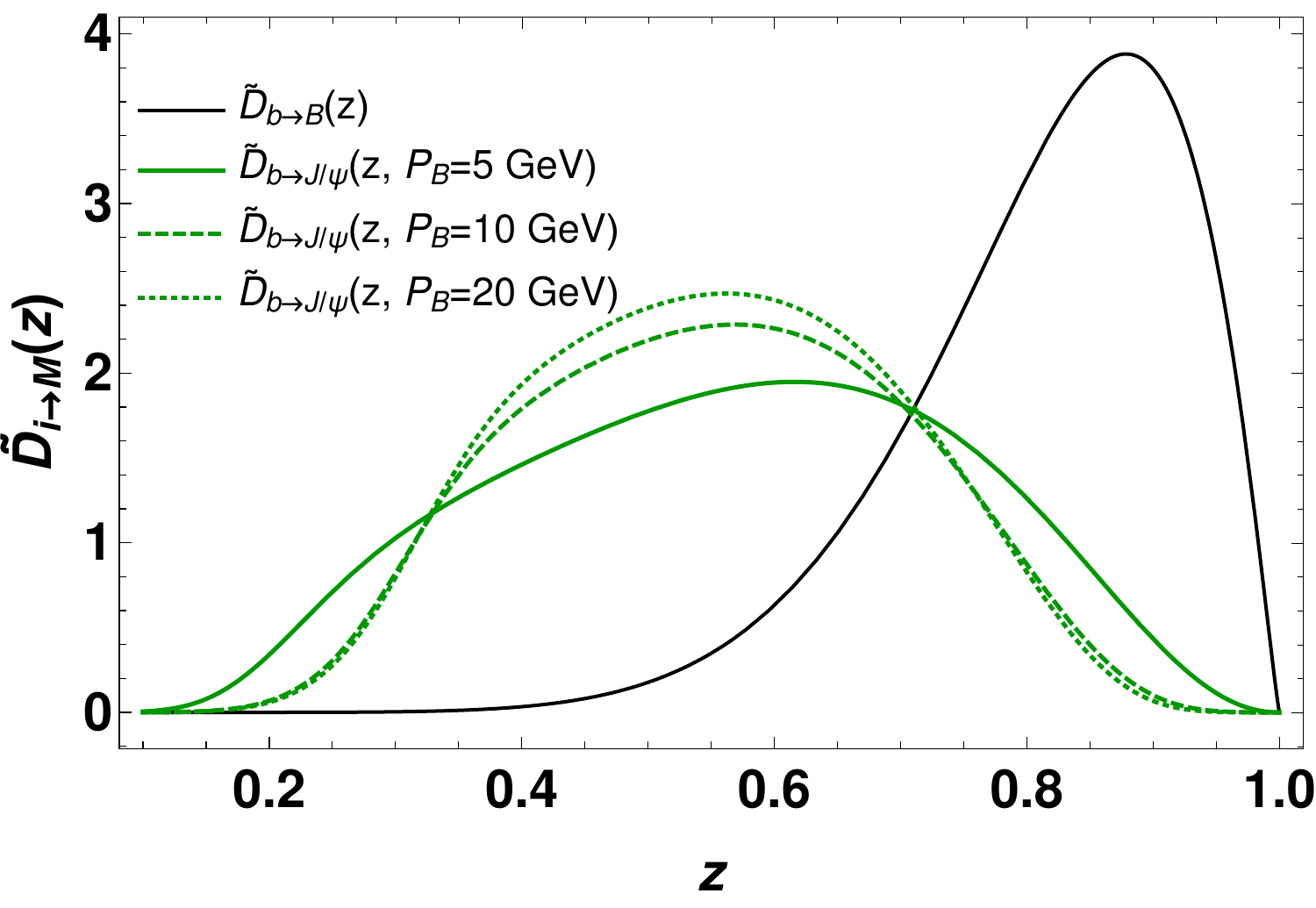}

\caption{\label{fig:fragFunction} The fragmentation function of $B$-quarks
and non-prompt $J/\psi$ mesons. To facilitate comparison of the shapes,
we normalized all the fragmentation functions to unity (so we use
the notation $\tilde{D}_{i\to M}$ instead of $D_{i\to M}$). The
normalization coefficients for $b\to B$ and $b\to J/\psi$ cases
differ by the branching fraction $Br_{B\to J/\psi}\approx0.8\,\%$. }
\label{DiagsMultiplicity1-1} 
\end{figure}

For the sake of completeness, in this appendix
we briefly summarize the fragmentation functions used in our evaluations.
Since the fragmentation functions are essentially nonperturbative
and cannot be evaluated from first principles, currently their parametrization
is extracted from the phenomenological fits of $e^{+}e^{-}$ annihilation
data. For the $B$-mesons the dominant contribution comes from the
fragmentation of $b$-quarks, and for the fragmentation function of
this process we used the parametrization from~\cite{Binnewies:1998vm}
\begin{equation}
D_{b\to B}\left(z,\,\mu_{0}\right)=N\,z^{\alpha}\left(1-z\right)^{\beta},\label{eq:Db1}
\end{equation}
where $N=56.4$, $\alpha=8.39$, $\beta=1.16$. The shape of parametrization~(\ref{eq:Db1})
is close to another widely used parametrization from~\cite{Peterson:1982ak}
\begin{align}
D_{b\to B}\left(z,\,\mu_{0}\right) & =\frac{N}{z\left(1-\frac{1}{z}-\frac{\epsilon}{1-z}\right)^{2}},\label{eq:Db2}\\
\epsilon & \approx0.0126
\end{align}
The production of non-prompt charmonia which stem from decays of the
$B$-mesons might also be described using a fragmentation function,
which is related to that of $B$-mesons as~\cite{Kniehl:1999vf}
\[
D_{b\to J/\psi}\left(z,\,\mu\right)=\int_{z}^{1}dx\,D_{b\to B}\left(\frac{x}{z},\,\mu^{2}\right)\times\frac{1}{\Gamma_{B}}\frac{d\Gamma}{dz}\left(z,\,P_{B}\right)
\]
where $\Gamma_{B}\equiv1/\tau_{B}$ is the total decay width of the
$B$-meson, and the function $d\Gamma\left(z,\,P_{B}\right)/dz$ was
evaluated in detail in~\cite{Kniehl:1999vf}. In the Figure~\ref{fig:fragFunction}
we compare the fragmentation functions $D_{b\to B}$ and $D_{b\to J/\psi}$.
These two functions differ by the branching fraction $Br_{B\to J/\psi}\approx0.8\,\%$,
and for this reason in order to facilitate comparison, we plotted
the fragmentation functions normalized to unity, $\tilde{D}(z)=D(z)/\int_{0}^{1}dz\,D(z)$.
As we can see, the distribution $D_{b\to J/\psi}$ is significantly
wider than $D_{b\to B}$ and has a peak near smaller values of $z\approx0.5$.

The $D$-mesons might be produced either from fragmentation of $c$-quarks
(prompt mechanism) or from $b$-quarks (non-prompt mechanism). The
fragmentation functions for both cases are available from~\cite{Kneesch:2007ey},
\begin{equation}
D_{i\to D}\left(z,\,\mu_{0}\right)=N_{i}\,z^{-\left(1+\gamma_{i}^{2}\right)}\left(1-z\right)^{a}\exp\left(-\gamma_{i}^{2}/z\right),\quad i=b,\,c\label{eq:Dc}
\end{equation}
with parameters given in the Table~\ref{tab:Parameters}. Though
the parameters for $D^{+}$ and $D^{0}$ in the table differ significantly,
their fragmentation functions have very similar shapes and differ
only by a factor of two in normalization.

\begin{table}
\begin{tabular}{|c|c|c|c|c|c|c|}
\hline 
 & $N_{c}$  & $a_{c}$  & $\gamma_{c}$  & $N_{b}$  & $a_{b}$  & $\gamma_{b}$\tabularnewline
\hline 
$D^{0}$  & $8.8\times10^{6}$  & $1.54$  & $3.58$  & $78.5$  & $5.76$  & $1.14$\tabularnewline
\hline 
$D^{+}$  & $5.67\times10^{5}$  & $1.16$  & $3.39$  & $185$  & $7.08$  & $1.42$\tabularnewline
\hline 
\end{tabular}\caption{\label{tab:Parameters}The values of parameters of $D$-meson fragmentation
function with parametrization~(\ref{eq:Dc}), as found in~\cite{Kneesch:2007ey}.}
\end{table}

\section{Parametrization for the matrix $\Omega_{ik}$}

\label{sec:SoftPomeron}In this appendix we briefly summarize the
parametrization of the soft pomeron scattering amplitude $\Omega_{ik}$
used in Section~\ref{subsec:LRGSF}. In the two-channel model it
is assumed that in addition to proton there is another diffractive
state $X$, which might be produced instead of proton in inelastic
processes (\emph{e.g}. single diffractive, double diffractive). The
matrix $\Omega_{ik}$ is thus a $2\times2$ matrix in the subspace
which includes a proton and the diffractive state $X$.

For our evaluations we used a parametrization from~\cite{Khoze:2018kna},
which has a form 
\begin{align}
\Omega_{ik}(b,\,s) & =\int\frac{d^{2}q}{4\pi^{2}}e^{iq\cdot b}\tilde{\Omega}_{ik}\left(t=-q^{2},\,s\right)\label{eq:KhozeFourier-1}\\
\tilde{\Omega}_{ik}(t,\,s) & =v_{i}F_{i}(t)F_{k}(t)\,\left(\frac{s}{s_{0}}\right)^{\alpha_{IP}-1},\\
F_{i}(t) & =\exp\left(b_{i}\left(c_{i}^{d_{i}}-\left(c_{i}-t\right)^{d_{i}}\right)\right),\\
s_{0} & \approx1\,{\rm GeV}^{2},\quad v_{1,2}=\sqrt{\sigma_{0}}(1\pm\lambda),\\
\sigma_{0} & \approx23\,{\rm mb},\,\,\lambda\approx0.56,\\
b_{1} & \approx10\,{\rm GeV}^{-2},\,b_{2}\approx4.9\,{\rm GeV}^{-2},\\
c_{1} & \approx0.233\,{\rm GeV}^{2},\,c_{2}\approx0.52\,{\rm GeV}^{2},\\
d_{1} & \approx0.462,\,\,d_{2}\approx0.47,\\
\alpha_{IP}(t) & \approx1.13+0.052\,t.
\end{align}
and has been fitted using recent LHC data on elastic, single diffractive
and double diffractive scattering. 

 \end{document}